\documentclass[final,5p,times,twocolumn,unknownkeysallowed]{elsarticle}

\usepackage{doi}
\usepackage{hyperref}
\hypersetup{
	colorlinks=true,        
	linkcolor=blue,         
	citecolor=cyan,         
}

\DeclareUnicodeCharacter{2212}{\textendash}
\usepackage{graphicx}
\usepackage{amssymb}
\usepackage{amsmath}
\usepackage{amsthm}
\usepackage{hyperref}
\usepackage{multirow}

\usepackage{dcolumn}
\usepackage{bm}
\usepackage{color}

\biboptions{comma,sort&compress}

\def\a{\alpha}
\def\r{\rho}
\def\s{\sigma}
\def\t{\tau}
\def\m{\mu}
\def\n{\nu}
\def\k{\kappa}
\def\th{\theta}
\def\g{\gamma}\def\G{\Gamma}
\def\L{\Lambda}\def\l{\lambda}
\def\D{\Delta}
\def\la{\langle}
\def\ra{\rangle}
\def\o{\omega}\def\O{\Omega}
\def\d{\delta}
\def\p{\partial}

\newcommand{\be}{\begin{equation}}
	\newcommand{\ee}{\end{equation}}
\newcommand{\bea}{\begin{eqnarray}}
	\newcommand{\eea}{\end{eqnarray}}

\def\half{\textstyle{\frac{1}{2}}}

\def\bdoc{\begin{document}}
	\def\edoc{\end{document}}
\def\beq{\begin{equation}}
	\def\eeq{\end{equation}}
\def\bea{\begin{eqnarray}}
	\def\eea{\end{eqnarray}}
\def\ben{\begin{enumerate}}
	\def\een{\end{enumerate}}
\def\la{\langle}
\def\ra{\rangle}
\def\a{\alpha}
\def\b{\beta}
\def\g{\gamma}
\def\G{\Gamma}
\def\d{\delta}
\def\D{\Delta}
\def\e{\epsilon}

\def\th{\theta}
\def\k{\kappa}
\def\l{\lambda}
\def\m{\mu}
\def\n{\nu}
\def\o{\omega}
\def\p{\pi}
\def\r{\rho}
\def\s{\sigma}
\def\t{\tau}
\def\L{{\cal L}}
\def\S{\Sigma }
\def\gsim{\; \raisebox{-.8ex}{$\stackrel{\textstyle >}{\sim}$}\;}
\def\lsim{\; \raisebox{-.8ex}{$\stackrel{\textstyle <}{\sim}$}\;}
\def\gtrsim{\gsim}
\def\lessim{\lsim}
\def\loc{{\rm local}}
\def\vm{v_{\rm max}}
\def\bh{\bar{h}}
\def\del{\partial}
\def\nab{\nabla}
\def\half{{\textstyle{\frac{1}{2}}}}
\def\fourth{{\textstyle{\frac{1}{4}}}}

\def\bD{{\bf D}}
\def\bE{{\bf E}}
\def\bF{{\bf F}}
\def\bB{{\bf B}}
\def\bP{{\bf P}}
\def\bV{{\bf v}}
\def\bv{{\bf v}}
\def\bx{{\bf x}}
\def\by{{\bf y}}
\def\bz{{\bf z}}
\def\ba{{\bf a}}
\def\bd{{\bf d}}
\def\bs{{\bf s}}
\def\bn{{\bf n}}
\def\bp{{\bf p}}

\def\O{\Omega}

\def\br{{\bf r}}
\def\bnab{{\bf \nab}}

\def\tE{\tilde{E}}
\def\tL{\tilde{L}}

\journal{Physics of the Dark Universe}

\begin{document}
	
	\begin{frontmatter}
		\allowdisplaybreaks
		\newcommand{\bq}{\begin{equation}}
			\newcommand{\eq}{\end{equation}}
		\newcommand{\bqn}{\begin{eqnarray}}
			\newcommand{\eqn}{\end{eqnarray}}
		\newcommand{\nb}{\nonumber}
		\newcommand{\lb}{\label}
		\newcommand{\f}{\frac}
		\newcommand{\mr}[1]{\textcolor{blue}{\bf Mirzabek: #1}}
		\newcommand{\PRL}{Phys. Rev. Lett.}
		\newcommand{\PLB}{Phys. Lett. B}
		\newcommand{\PRD}{Phys. Rev. D}
		\newcommand{\CQG}{Class. Quantum Grav.}
		\newcommand{\JCAP}{J. Cosmol. Astropart. Phys.}
		\newcommand{\JHEP}{J. High. Energy. Phys.}
		\newcommand{\red}{\textcolor{red}}
		\def\cred#1{\textcolor{red}{#1}}
		\def\cblue#1{\textcolor{blue}{#1}}
		\def\colive#1{\textcolor{olive}{#1}}
		\title{Particle dynamics and the accretion disk around a Self-dual Black Hole immersed in a magnetic field in Loop Quantum Gravity
		}
		\author[mainaddress1]{Uktamjon Uktamov}
		\ead{uktam.uktamov11@gmail.com}
		\author[mainaddress1,mainaddress2,shakhrisabz]{Mirzabek Alloqulov\cortext[cor2]{Corresponding author}\corref{cor2}
		}
		\ead{malloqulov@gmail.com}
		\author[mainaddress3,mainaddress1,mainaddress2,caspian]{Sanjar Shaymatov}
		\ead{sanjar@astrin.uz}
		\author[mainaddress3,mainaddress4]{Tao Zhu}
		\ead{zhut05@zjut.edu.cn}
		\author[mainaddress1]{Bobomurat Ahmedov}
		\ead{ahmedov@astrin.uz}
		
		\address[mainaddress1]{Institute of Fundamental and Applied Research, National Research University TIIAME, Kori Niyoziy 39, Tashkent 100000, Uzbekistan}
		\address[mainaddress2]{Tashkent University of Applied Sciences, Gavhar Str. 1, Tashkent 100149, Uzbekistan}
		\address[shakhrisabz]{Shahrisabz State Pedagogical Institute, Shahrisabz Str. 10, Shahrisabz 181301, Uzbekistan}
		\address[mainaddress3]{Institute for Theoretical Physics and Cosmology, Zhejiang University of Technology, Hangzhou, 310032, China}
		\address[caspian]{Western Caspian University, Baku AZ1001, Azerbaijan}
		\address[mainaddress4]{United Center for Gravitational Wave Physics (UCGWP), Zhejiang University of Technology, Hangzhou, 310032, China}

		\date{\today}
		\begin{abstract}
			In this paper, we study the motion of magnetic dipoles and electrically charged particles in the vicinity of a self-dual black hole in Loop Quantum Gravity (LQG) immersed in an external asymptotically uniform magnetic field. We explore the effects of the quantum correction parameter and electromagnetic interactions on the particle geodesics. We derive the field equations and determine the electromagnetic four-vector potential for the case of a self-dual black hole in LQG. We investigate the innermost stable circular orbits (ISCOs) for both magnetic dipoles and electrically charged particles in detail, demonstrating that the quantum correction parameter significantly influences on the ISCO radius, causing it to shrink. Additionally, we show that the ISCO radius of magnetic dipoles is greater than that of electrically charged particles due to the magnetic field interaction. We investigate the ISCO parameters (i.e., $r_{ISCO}$, $l_{ISCO}$, $\mathcal{E}_{ISCO}$, $v_{ISCO}$, and $\Omega_{ISCO}$) for magnetic dipoles and electrically charged particles, providing detailed values. Furthermore, we examine the trajectories of charged particles under various scenarios resulting from the quantum correction parameter $P$. Finally, analyzing the ISCO parameters that define the inner edge of the accretion disk, we explore the accretion disk around a self-dual black hole in LQG. We delve into the electromagnetic radiation flux, temperature, and differential luminosity as radiation properties of the accretion disk in detail. We show that the quantum correction parameter shifts the profile of the electromagnetic flux and accretion disk temperature towards the central object, leading to a slight increase in these quantities.
			
		\end{abstract}
	\end{frontmatter}
	
	
	\section{Introduction}
	\renewcommand{\theequation}{1.\arabic{equation}} \setcounter{equation}{0}
	
	Even though general relativity (GR) has been considered the most successful theory of gravity, it has faced notable theoretical and observational challenges. From a theoretical point of view, GR's lack of incorporation of quantum principles hinders its unification with quantum mechanics, a significant unresolved issue  \cite{Adler10,Ng03MPLA}. Furthermore, GR predicts singularities, points where known physics breaks down, both at the universe's inception \cite{Borde03PRL,Borde94PRL} and within black hole spacetimes \cite{Hawking73}. Observationally, GR struggles to explain phenomena such as dark matter and dark energy. All these issues indicate that the classical GR might need to be modified. In particular, the spacetime singularities ought to be resolved after quantum gravitational effects are taken into account. These challenges motivate the exploration of modified gravity theories and quantum gravity as potential solutions. Rigorous testing of these modified theories is crucial to determine which, if any, can ultimately provide a comprehensive and accurate model of gravity. 
	
	Loop quantum gravity (LQG) presents an elegant solution to classical Big Bang and black hole singularities. Recently, a regular static spacetime metric known as the self-dual spacetime has been derived within a mini-superspace framework using the polymerization procedure in LQG \cite{Modesto10}. This spacetime demonstrates regularity and lacks any spacetime curvature singularity. The polymerization procedure in the quantization of the black hole spacetime introduces two intrinsic parameters: the minimal area and the Barbero-Immirzi parameter, which control the quantum effects on the self-dual spacetime. Furthermore, it has been shown that this spacetime exhibits self-duality, aligning with T-duality principles \cite{Modesto09, Sahu15}. This is also the reason that we call it the self-dual spacetime. In recent years, extensive research has focused on the quantum extensions of black holes within the LQG framework, with notable references including \cite{Ashtekar18PRL, Ashtekar18, Bojowald18, Alesci19PLB, Assanioussi20, Bodendorfer:2019nvy, Bodendorfer:2019jay}. For a more comprehensive understanding, readers are encouraged to explore review articles such as \cite{Perez17, Barrau18, Rovelli18, Ashtekar20, Gan20}.

	The quantum effects on the self-dual spacetime are controlled by the two intrinsic parameters: the minimal area and the Barbero-Immirzi parameter. This raises the intriguing question of whether these quantum effects can leave any observational signatures for the current and/or forthcoming experiments so that the quantum effects can be tested or constrained directly. Such considerations have resulted in a flourish of studies during the past decades from different kinds of experiments and observations \cite{Alesci14, Chen11, Barrau14, Dasgupta13,Hossenfelder12:arXiv,Sahu15,Cruz19,Moulin19a,Moulin19b,Cruz20,Santos21,Liu20,Zhu20}.  For instance, a detailed exploration in \cite{Liu20} elaborates on how the quantum effects of LQG can impact the shadow of a rotating black hole, offering comparisons with the latest observations by the Event Horizon Telescope (EHT) of the supermassive black hole M87*. Furthermore, the gravitational lensing in the self-dual spacetime has been under scrutiny, with the polymeric function from LQG being constrained using data from geodetic very-long-baseline interferometry measurements of the solar gravitational deflection of radio waves \cite{Sahu15}. Recent studies, like those in \cite{Zhu20}, have extended this investigation to the solar system, deriving observational constraints on the polymeric function $P$ of LQG.  It is interesting to note that the phenomenological study of the self-dual black hole and other types of loop quantum black holes has also been extensively explored, see \cite{Liu21PRD,Daghigh21,Bouhmadi-Lopez20JCAP,Fu22,Brahma21PRL, Liu:2023vfh, Tu:2023xab, Yang:2023gas, Yan:2023vdg, Albuquerque:2023lhm, Chen:2024sbc, Jiang:2023img, Jiang:2024vgn, Chen:2023bao, Yan:2022fkr, Gingrich:2023fxu, Liu:2024qci, Liu:2022qiz} and references therein.
	
	In this paper, we consider a self-dual black hole in LQG in the presence of an external magnetic field. To understand the nature of spacetime geometry we analyze the effects of a new quantum correction parameter and existing fields in the surrounding environment. For that, we investigate the motion of particles with magnetic dipole moment and electrically charged particles around the aforementioned spacetime geometry. Additionally, we examine the properties of the accretion disk around the black hole to provide valuable insights into the properties of a self-dual black hole in LQG and its geometry.
	
	The important point to note is that the solution of either GR or alternative theories of gravity can be tested through astrophysical processes, such as X-ray data \cite{Bambi12a,Bambi16b} and the accretion disk \cite{Abramowicz13} with recent observations \cite{Fender04mnrs,Auchettl17ApJ,IceCube17b} around astrophysical compact objects, playing a pivotal role as very useful tests. On the other hand, as one possible attempt, it is also valuable to analyze the motion of test particles around compact objects to understand more deeply their unique aspects. The particle motion can also be used to model the accretion disk~\cite{Bambi17e,Chandrasekhar98}. Therefore, the aforementioned tests may come into play to reveal unknown aspects of astrophysical compact objects and to provide some departures from their expected behaviors. It is to be emphasized that, from an astrophysical viewpoint, the presence of external magnetic fields can play a pivotal role in determining the motion of charged particles in the vicinity of compact objects. The effects of electromagnetic fields have been widely considered and examined on test particles with an electric charge and with a nonzero magnetic dipole moment around black holes in different gravity theories~\cite{deFelice03,deFelice04,Vrba20,Shaymatov21d,Frolov11,Frolov12,Hussain17,Kolos17,Narzilloev20a,Narzilloev20b,Kovar14,Shaymatov15,Duztas-Jamil20,Shaymatov20egb,Shaymatov23GRG,Shaymatov21c,Shaymatov21pdu,Shaymatov24MR,Khan:2024jez,Shaymatov24MP}. 
	
	In an astrophysical scenario, a black hole does not have its
	own magnetic field due to gravitational collapse, which decays at least 
	with $t^{−1}$ in the quasistationary approach (see, for example,  \cite{Ginzburg1964,Anderson70}). 
	However, one can
	assume existence of an external test magnetic field that can be produced by nearby objects, e.g., accretion disks existing around rotating black holes \cite{Wald74}, neutron stars \cite{deFelice03,Ginzburg1964,Rezzolla01}, and magnetars \cite{Morozova12,Kovar14_Magnetar}. The magnetic field existing in the surrounding environment was estimated to be of the order of $B_1\sim 10^{8}~\rm{G}$ for stellar and $B_2\sim 10^{4}~\rm{G}$ for supermassive black holes, respectively (see, e.g., ~\cite{Piotrovich10,Baczko16,Daly:APJ:2019:}). It does not significantly distort the spacetime geometry, so it can be considered a test field that satisfies $B_{1,2}\ll B_{max} \sim 10^{19}({M_\odot}/{M})~ \rm{G}$. 
	However, recent observational findings from EHT on M87 regarding the magnetic field strength around black holes reported that the magnetic field is of the order of $B\sim 33.1 \pm 0.9~\rm{G}$ in the corona for the binary black hole system $V404$ Cygni~\cite{Dallilar2018}. Additionally, EHT collaborations announced in their report that the average magnetic field strength is of the order of $B\sim 1 - 30~\rm{G}$ in the emission region around the supermassive black hole of the M87 galaxy~\cite{MF:2021ApJ,Narayan2021ApJ}.  In this paper, we explore properties of  the self-dual black hole in LQG immersed
	in an external magnetic field, which is asymptotically uniform at large
	distances and oriented along the $z$-axis. However, it should be emphasized that, in a realistic astrophysical scenario, the magnetic field can have a complex configuration in the close vicinity of a black hole, especially very close to its horizon, but it can be approximated as a uniform field at a specific distance, particularly in a local part of space. This is a well-accepted approximation for magnetic fields in the surrounding environment of black holes. For example, the magnetic field generated by a current loop in the accretion disk around the black hole can be approximated to be uniform provided that the black hole size is much smaller compared to the disk size (see, e.g., \cite{Petterson1974,Frolov10}). Additionally, the magnetic field can be considered as a nearly uniform field when the black hole is located at a sufficient distance away from the equatorial plane of a magnetar \cite{Kovar14_Magnetar,Stuchlik16EPJC}. These investigations suggest that the magnetic field can be regarded as an asymptotically uniform magnetic field at a relatively far distance.

	This paper is organized as follows: In Sec. \ref{Sec:LQG}, we briefly discuss a black hole spacetime in LQG and the surrounding magnetic field properties. In Sec.~\ref{Sec:mag_dip}, we study the motion of particles with a magnetic dipole moment, which is followed by the study of dynamics of electrically charged particles around a self-dual black hole immersed in an external uniform magnetic field in Sec.~\ref{Sec:elec_par}. We further consider the accretion disk around the black hole and examine its radiative properties, e.g., the electromagnetic
	radiation flux, the temperature and the differential luminosity of the accretion disk in Sec~\ref{Sec:accr_disk}. We end up with a conclusion in Sec. \ref{Sec:Con}. In  \ref{A}, we consider the solution of field equations associated with the magnetic field existing in the environment surrounding a self-dual black hole in LQG. We use $(–, +, +, +)$ for the spacetime metric and the system of units $c=G=M=1$ throughout the paper.

	\section{Loop quantum black hole immersed in the external uniform magnetic field}\label{Sec:LQG}
	
	We begin with an introduction to the quantum corrected Schwarzschild spacetime in LQG proposed in Ref.~\cite{Modesto10}, commonly referred to as the self-dual spacetime. In effective models of black holes within LQG, a polymer-like quantization method is typically employed. This approach involves replacing the canonical variables $(b, c)$ in the phase space of black hole spacetime with their regularized counterparts: $b \to \frac{\sin(\delta_b b)}{\delta_b}$ and $c \to \frac{\sin(\delta_c c)}{\delta_c}$. The parameters $\delta_b$ and $\delta_c$ are quantum polymeric parameters that determine the scales at which quantum effects manifest in LQG. In the limit where $\delta_b, \delta_c \to 0$, the classical physics is recovered. However, a complete theory of quantum gravity is not yet available, and thus, a comprehensive framework for choosing $\delta_b$ and $\delta_c$ remains unclear. In this paper, we explore the self-dual spacetime arising from a specific choice that treats $\delta_b$ and $\delta_c$ as constants, known as the $\mu_0$-scheme in LQG. Ref. \cite{Modesto10} demonstrates that identifying the minimal area in the solution with the minimum area $A_{\text{min}}$ of LQG allows for one of the free parameters to be reduced. Consequently, in the self-dual spacetime, the LQG effects are characterized by two parameters: the minimal area $A_{\text{min}}$ and the polymeric parameter $\delta$ (where $\delta_b$ is denoted as $\delta$ hereafter).

	The effective metric of the self-dual black hole in LQG is given by \cite{Modesto10}
	\begin{equation}
		{d\tau}^2=-A(r){dt}^2+\frac{{dr}^2}{B(r)}+C(r)({d\theta}^2+\sin^2{\theta}{d\phi}^2),
		\label{self-dual Schwarzschild metric}
	\end{equation}
	where $A(r)$, $B(r)$, $C(r)$ are given by
	\begin{equation}
		\begin{split}
			&A(r)=-g_{tt}=\frac{(r-r_{+})(r-r_{-})(r+r_{*})^2}{r^4+a_{0}^{2}},\\
			&B(r)=\frac{1}{g_{rr}}=\frac{(r-r_{+})(r-r_{-})r^4}{(r+r_{*})^2(r^4+a_{0}^{2})},\\
			&C(r)=\frac{g_{\phi\phi}}{\sin^2\theta}=r^2+\frac{a_{0}^{2}}{r^2}\, ,
		\end{split}
		\label{ABC}
	\end{equation}
	\noindent{where $r_{+}=2M/(1+P)^2$ and $r_{-}=2MP^2/(1+P)^2$ refers to the black hole's two horizons, while $r_{*}=\sqrt{r_{+}r_{-}}=2MP/(1+P)^2$ with ADM mass $M$ of the black hole. Here, we only restrict ourselves to the outer horizon. We note here that the function $P$ stemming from quantum correction is given by
		\begin{equation}
			P\equiv\frac{\sqrt{1+\varepsilon^2}-1}{\sqrt{1+\varepsilon^2}+1}.
			\label{P}
		\end{equation}
		
		\begin{figure*}[!htb]
			\centering
			\includegraphics[width=0.3\textwidth]{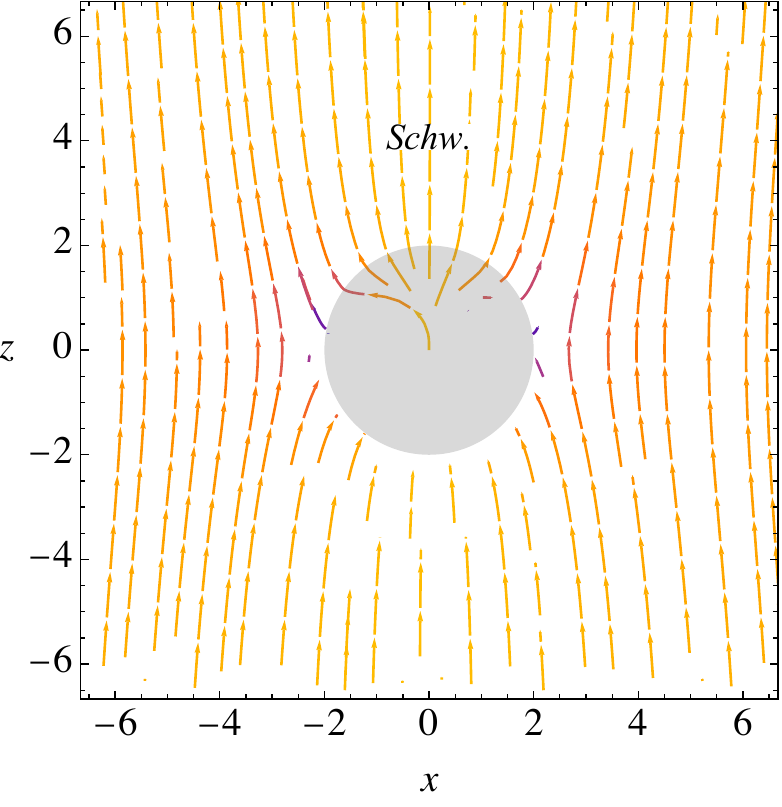}
			\includegraphics[width=0.3\textwidth]{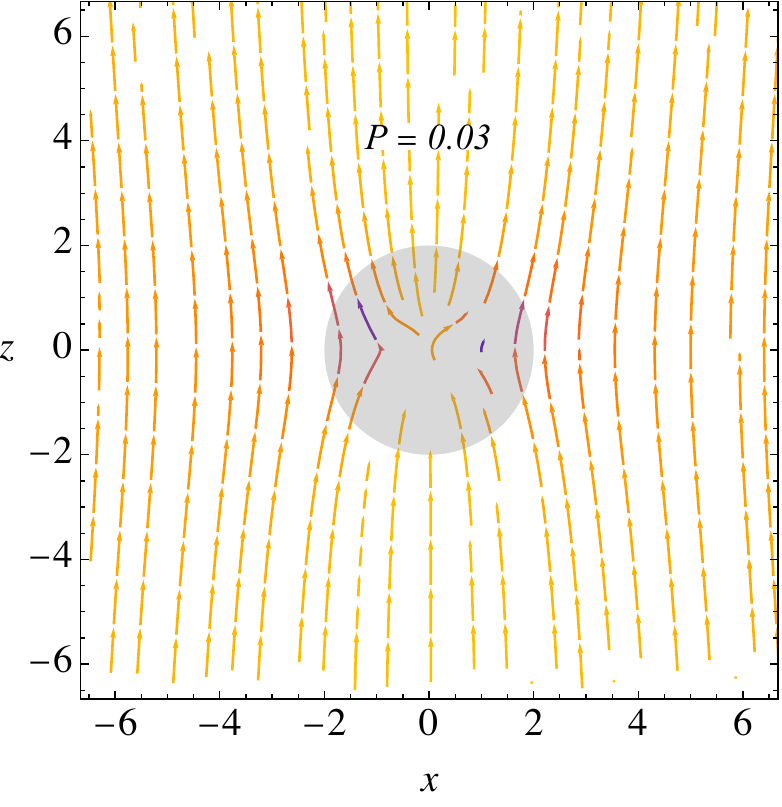}
			\includegraphics[width=0.3\textwidth]{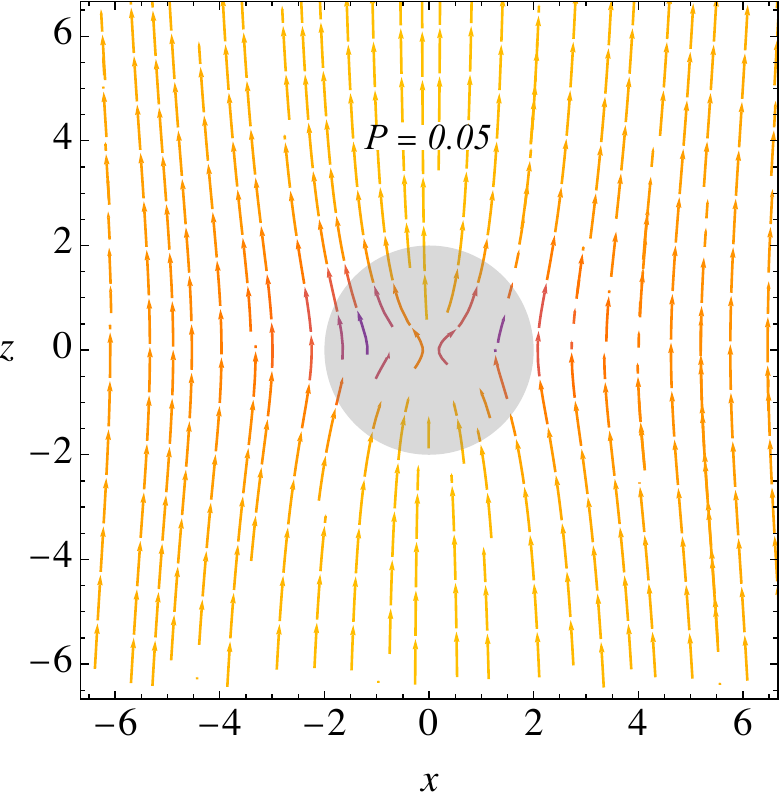} %
			
			\caption{\label{Fields}
				The plot shows the magnetic field lines in the vicinity of the self-dual black hole in LQG for various combinations of the quantum correction parameter $P$, referred to as $\epsilon=0.3571$ and $\epsilon=0.4707$, respectively.}
		\end{figure*}
		
		\noindent{Here, $\varepsilon=\gamma \delta$ denotes a product of the Immirzi parameter $\gamma$ and the polymeric parameter $\delta$ which is introduced to define the paths one integrates the connection along to define the holonomies in the quantum corrected Hamiltonian constraint in the polymerization procedure in LQG, satisfying $\varepsilon<1$.} The point to be noted is that we further focus on $P$ only to explore the quantum effects resulting from LQG. The $a_0$ parameter is defined by the equation $a_0=\frac{A_{min}}{8 \pi}$ with $A_{min} \approx 4 \pi \gamma \sqrt{3}l^2_{Pl}$ refers to the minimum area gap of LQG. Here, $l_{Pl}$ is the Planck length. The Planck length can be considered negligible. Therefore we consider  $a_{0}=0$. Here our main objective is to examine a charged particle motion in the close vicinity of the LQG black hole in the presence of an external magnetic field that refers to as asymptotically uniform with the strength $B_0$. To that end, we further need to find the four potential of the electromagnetic field, $A^{\mu}$. Following \cite{Wald74,2016EPJC...76..414A}, we assume the magnetic field to be uniform and oriented along the axis of symmetry of the LQG black hole due to its asymptotic properties. In doing so, one can write the vector potential's ansatz for a non-rotating black hole as follows:
		\begin{eqnarray}\label{1.4}
			A^\mu=c_t(r,B_0)\xi_t^\mu+\Big[\frac{B_0}{2}+c_\phi(r,B_0)\Big]\xi_\phi^\mu\, ,  
		\end{eqnarray}
		where $\xi_t^\mu=(1,0,0,0)$ and $\xi_\phi^\mu=(0,0,0,1)$ refer to the time-like and space-like killing vectors. The explicit form of $A^{\mu}$ can be determined as a solution of the source-free Maxwell equations \cite{2016EPJC...76..414A}. For a non-rotating vacuum black hole the Maxwell field equations can be defined by $F^{\phi\nu}_{;\nu}=0$ with the electromagnetic field tensor $F_{\mu\nu}=\partial_\mu A_\nu-\partial_\nu A_\mu$. One can then write the Maxwell field equation for $A^{\mu}$ in the LQG black hole spacetime as follows:
		\begin{align} \label{Eq:maxwell_field}
			A_{;\nu}^{\phi;\nu}-R_\nu^\phi A^\nu = 0 \, .
		\end{align}
		Similarly, the same form is satisfied for Killing vector $\xi$, and it is written as
		\begin{align}\label{Eq:killing_field}
			\xi_{;\nu}^{\mu;\nu}-R_\nu^\mu \xi^\nu = 0 \, ,
		\end{align}
		with the Ricci tensor $R_{\mu\nu}$. Eq.~(\ref{Eq:killing_field}) takes the simpler form in the case of source-free satisfying the condition with Ricci tensor vanished, i.e., $\Box \xi^\mu=0$ that can then be extended the same form of the Maxwell equation, $\Box A^\mu=0$. Based on the equations underlined above with Eq.~(\ref{1.4}) one can have the following differential equation:
		\begin{eqnarray}\label{1.5} 
			&C_\phi''r^2 (r-r_+) (r-r_-)+C_\phi'r\left[2 (r_+ r_{-}-2r^2)\right.\nonumber\\&-\left.3r( r_++r_-)\right]
			-2 C_\phi (r_+ r_-+r_* (r_*+2 r))=0,
		\end{eqnarray}
		where we have denoted $C_\phi=\Big[\frac{B_0}{r}+c_\phi(r,B_0)\Big]$. Note that $^{\prime}$ and $^{\prime\prime}$ denote the first and second derivatives with respect to $r$. We provide details in relation to solution of the above equation in \ref{A}.  Here, the solution for the four potential is given by 
		\begin{eqnarray}\label{1.6}
			A^\mu=\frac{B_0}{2}  (\mathcal{B})^{\mathcal{A}}\,\,_2F_1\left(\mathcal{A},\mathcal{A}+3;2\mathcal{A}+3;\mathcal{B}\right)\xi_\phi^\mu.
		\end{eqnarray}
		Here, $_2F_1(a,b;c;z)=\Sigma_{n=0}^{\infty}\frac{(a)_n(b)_n}{(c)_n}\frac{z^n}{n!}$ is the Hypergeometric function, while $\mathcal{A}=\sqrt{\frac{1-2P}{1+2P}}-1$ and $\mathcal{B}=\frac{r(1+4P)}{2M(1+2P)}$ refer to new variables, as introduced in \ref{A}. It is obvious that for $P=0$ the four potential is reduced to the Schwarzschild case, i.e., $A_\phi=\frac{1}{2}B_0r^2\sin^2{\theta}$. In Fig.~\ref{Fields}, we demonstrate the magnetic field lines in the vicinity of the self-dual black hole in LQG. From Fig.~\ref{Fields}, one can observe that the quantum correction parameter $P$ can influence the magnetic field lines in the close vicinity of the self-dual black hole, compared to the Schwarzschild black hole case. On the basis of the four potential Eq.~(\ref{1.6}), it is straightforward to obtain non-vanishing components of electromagnetic tensors, and they are given as follows:
		\begin{eqnarray}
			F_{r\phi}&=&\frac{(\mathcal{B})^{\mathcal{A}}}{2} B_0 r \sin ^2{\theta} \nonumber\\&\times&\Big[\frac{\mathcal{B} \mathcal{C} r \, _2F_1\left(\mathcal{A}+1,\mathcal{A}+4;2(\mathcal{A}+2);\mathcal{B}\right)}{\Xi}+ \nonumber\\
			&+&\left(\mathcal{A}+2\right)_2F_1\left(\mathcal{A},\mathcal{A}+3;2\mathcal{A}+3;\mathcal{B}\right)\Big],
			\label{Elec.1}
		\end{eqnarray}
		\begin{equation}
			F_{\theta\phi}=\frac{(\mathcal{B})^{\mathcal{A}}}{2} B_0r^2\, \sin {2 \theta} \times
			\,_2F_1\left(\mathcal{A},\mathcal{A}+3;2\mathcal{A}+3;\mathcal{B}\right) . 
			\label{Elect.2}
		\end{equation}
		It is to be emphasized here that we denote new variables, such as $\mathcal{C}=\sqrt{1-4P^2}-6P-1$ and $\Xi=2\sqrt{1-4P^2}+2P+1$. In doing so, we can then unable to find the magnetic field components using $B^\beta=-\frac{1}{\sqrt{-g}}\epsilon^{\alpha\beta\sigma\gamma}F_{\sigma\gamma}u_\alpha$. Here, $g$ is the determinant of the metric Eq.~(\ref{self-dual Schwarzschild metric}) and $\epsilon^{\alpha\beta\sigma\gamma}$ the pseudo-tensorial form of the Levi-Civita
		symbol and $u_\alpha$ the four velocity of the observer. After that the orthonormal component of the magnetic field can be obtained as
		\begin{eqnarray}
			B^{\Hat{\theta}}=\frac{F_{r\phi}}{\sqrt{g_{rr}g_{\phi\phi}}}.
		\end{eqnarray}
		Here we assume that a distant observer is located at zero angular momentum observer (ZAMO). Therefore, the normalization condition, $u_\alpha u^\alpha=-1$, allows to find $u_t=\frac{1}{\sqrt{-g^{tt}}}$. 
		
		We further investigate the particle dynamics for two different cases around the black hole in LQG in the presence of an external magnetic field. This is what we intend to examine in the next section.

		\section{The motion of particles with nonzero magnetic dipole moment near a self-dual black hole immersed in an external magnetic field in LQG}\label{Sec:mag_dip}
		
		We now consider the motion of particles with a magnetic dipole moment around the black hole in LQG immersed in an external magnetic field. To that end we write the Hamilton-Jacobi equation for the particle interacting with the magnetic field, i.e., it is given by (\cite{deFelice03,PhysRevD.110.084084}):
		\begin{eqnarray}\label{Ham.}
			g^{\mu\nu}\frac{\partial S}{\partial x^\mu}\frac{\partial S}{\partial x^\nu}=-m^2\left(1-\frac{\mathcal{U}}{m}\right)\, ,
		\end{eqnarray}
		where $\mathcal{U}=D^{\mu\nu}F_{\mu\nu}$ refers to the magnetic interaction as a product of the polarization tensor $D^{\mu\nu}$ and electromagnetic tensor $F_{\mu\nu}$, which can also be defined by $\mathcal{U}=\mu_{\Hat{\alpha}}B^{\Hat{\alpha}}$. Following to \cite{Preti04}, the Lagrangian of the system for a magnetized particle motion is written as follows
		\begin{eqnarray}\label{Lagrangian}
			\mathcal{L}=\frac{1}{2}(m+\mathcal{U})g_{\mu\nu}u^{\mu}u^\nu-\frac{1}{2}\mathcal{U}.
		\end{eqnarray}
		From the Lagrangian (\ref{Lagrangian}), one is able to find the four-momentum as
		\begin{eqnarray}\label{momentum}
			p_{\mu}=\frac{\partial\mathcal{L}}{\partial\Dot{x}^\mu}=\left(m+U\right)g_{\mu\nu}u^\nu\, .
		\end{eqnarray}
		Based on the separable Hamilton-Jacobi action, $S=-Et+L\phi+S_r+S_\theta$, the equation for the magnetized particle can be obtained as 
		\begin{eqnarray}\nonumber
			g_{tt}\Big[\frac{l^2}{r^2\sin^2{\theta}}+1-\frac{\mathcal{U}}{m}+g_{rr}\left(1+\frac{\mathcal{U}}{m}\right)^2\Dot{r}^2\Big]+\mathcal{E}^2 + \\ \label{tr.mag}+g_{tt}g_{\theta\theta}\left(1+\frac{\mathcal{U}}{m}\right)^2\Dot{\theta}^2=0\, .
		\end{eqnarray}
		
		For further analysis, we shall restrict ourselves to the equatorial plane (i.e., $\theta=\pi/2$) and then assume that the magnetic dipole of the particles is perpendicular to the equatorial plane, thus allowing us to have the components of the particle's magnetic moment as $\mu^{\Hat{\alpha}}=(0,0,\mu,0)$ (\cite{deFelice03}). That is, the particle's magnetic moment is assumed to be oriented with the magnetic field aligned along the $z$-axis, typically perpendicular to the equatorial plane. This orientation allows the dipole moment to interact with the magnetic field by surviving its $\mu^{\Hat{\theta}}$ component, following the energy minimization principle. This choice not only simplifies our calculations but also enhances our qualitative understanding of magnetic interactions under the gravitational field of the self-dual black hole in LQG. This, in turn, allows one to have $\mathcal{U}=\frac{\mu F_{r\phi}}{\sqrt{g_{rr}g_{\phi\phi}}}$ for the magnetic interaction. 
		For the magnetized particle motion we then rewrite the Hamilton-Jacobi action at the equatorial plane as 
		\begin{eqnarray}\label{action}
			S=-Et+L\phi+S_r\, ,
		\end{eqnarray}
		where we denote the total energy and angular momentum of the magnetized particle by $p_t=-E$ and $p_\phi=L$, respectively. Taking all together, the motion of the magnetized particles around the LQG black hole takes the following forms 
		\begin{eqnarray}\nonumber
			\Dot{t}=-\frac{\mathcal{E}}{(1+\frac{\mathcal{U}}{m})g_{tt}}\, , 
		\end{eqnarray}
		\begin{eqnarray}\nonumber
			\Dot{\phi}=\frac{l}{(1+\frac{\mathcal{U}}{m})g_{\phi\phi}}\, ,
		\end{eqnarray}
		\begin{eqnarray}\label{motion}
			\Dot{r}^2=\frac{1}{-g_{rr}g_{tt}}(\mathcal{E}^2-V_{eff})\, ,
		\end{eqnarray}
		where $\mathcal{E}=E/m$ and $l=L/m$ refer to the specific energy and angular momentum of particles, respectively. From Eq.~(\ref{motion}) the effective potential for radial motion can be obtained as 
		\begin{eqnarray}\label{effective}
			V_{eff}=A(r)\left(1+\frac{l^2}{r^2}-\frac{\mathcal{U}}{m}\right)\, ,
		\end{eqnarray}
		which includes the dimensionless the magnetic interaction/coupling parameter 
		\begin{eqnarray}
			\beta=\frac{\mu B_0}{m}\, ,
		\end{eqnarray}
		which characterizes the magnetic field strength on the particles with a magnetic dipole or simply represents the interaction between external uniform magnetic field and these magnetized particles.
		\begin{figure*}[!htb]
			\centering
			\includegraphics[width=0.45\textwidth]{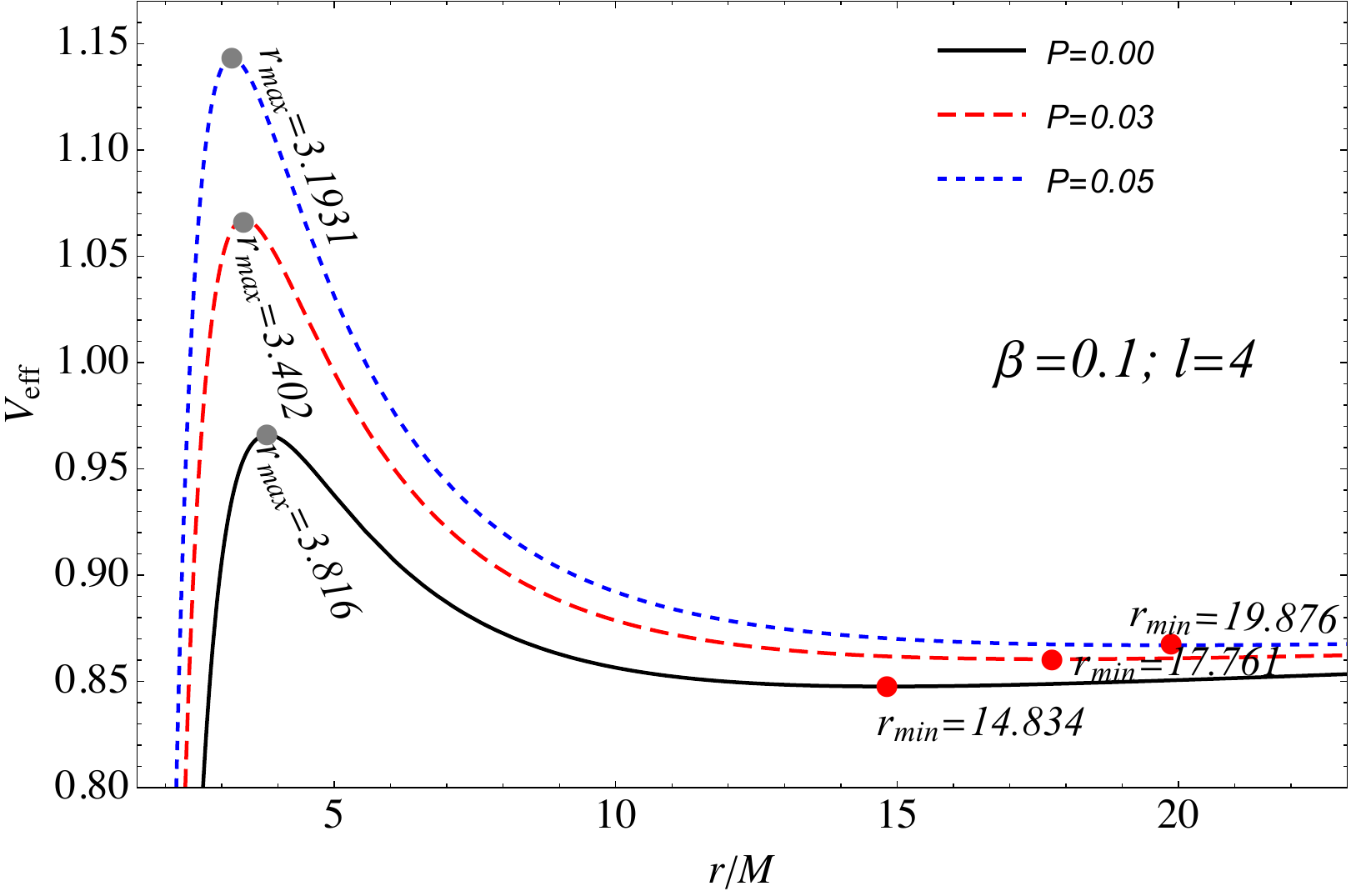}
			\includegraphics[width=0.445\textwidth]{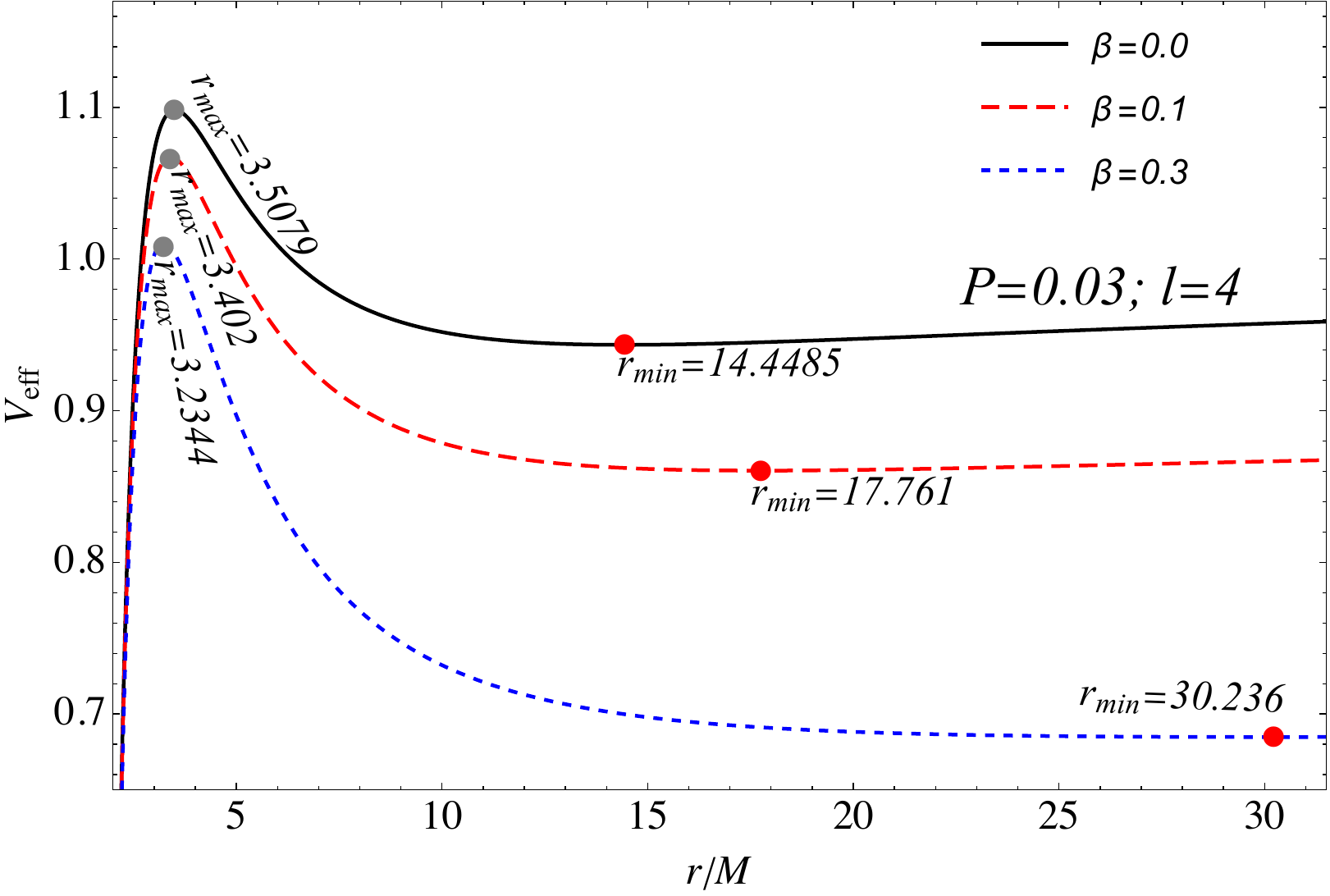}
			\caption{\label{effective} Radial profile of the effective potential for the particles with a nonzero magnetic dipole momentum around the self-dual black hole in the presence of an external asymptotically 
				uniform magnetic field for different selected  combinations of the quantum correction parameter $P$ (left panel) and the magnetic interaction parameter $\beta$ (right panel). One can observe 
				that the minima ($r_{min}$) and the maxima ($r_{max}$) correspond to the stable and unstable circular orbits, respectively.}
		\end{figure*}
		\begin{figure*}[!htb]
			\centering
			\includegraphics[width=0.32\textwidth]{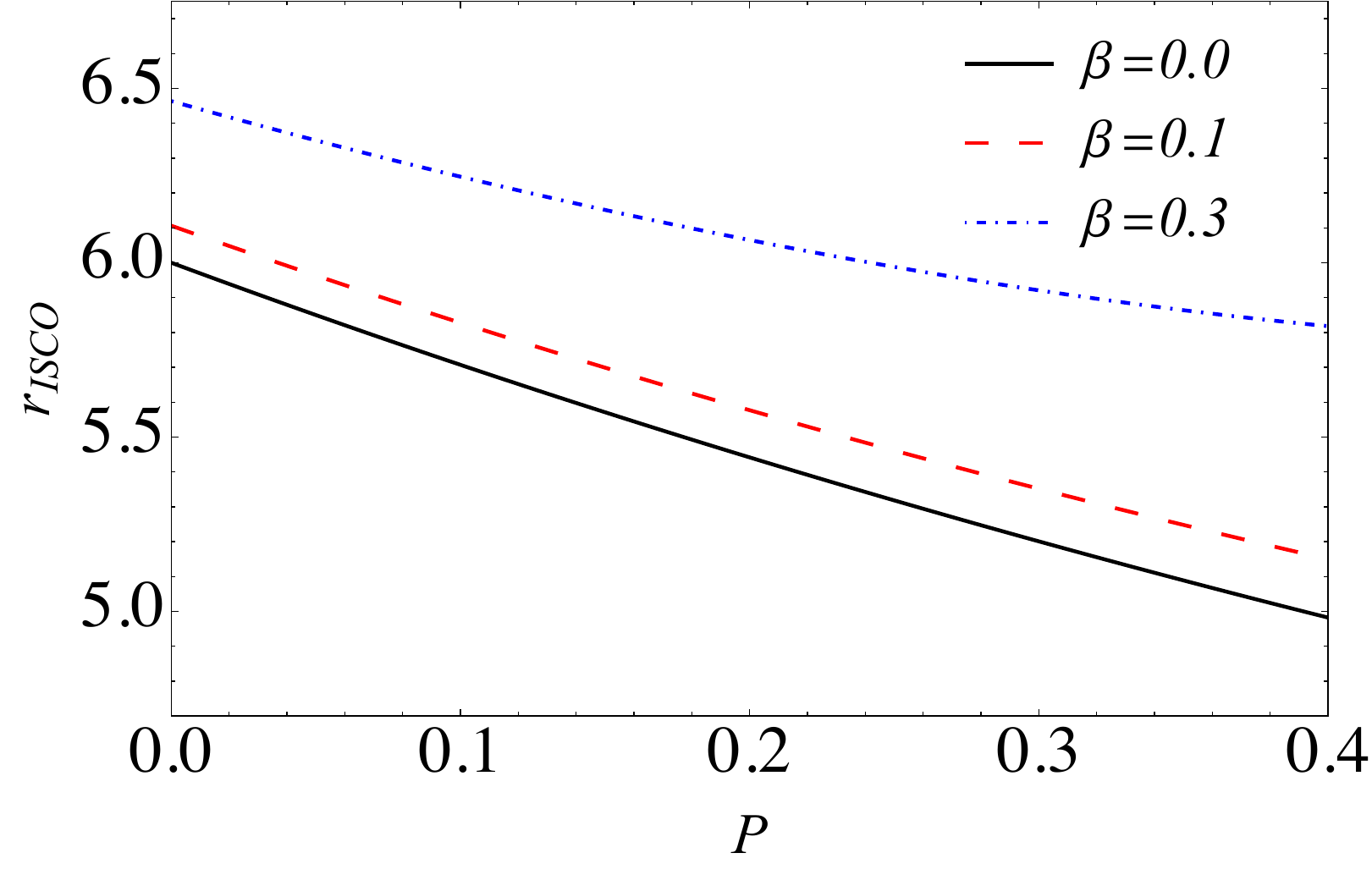}
			\includegraphics[width=0.32\textwidth]{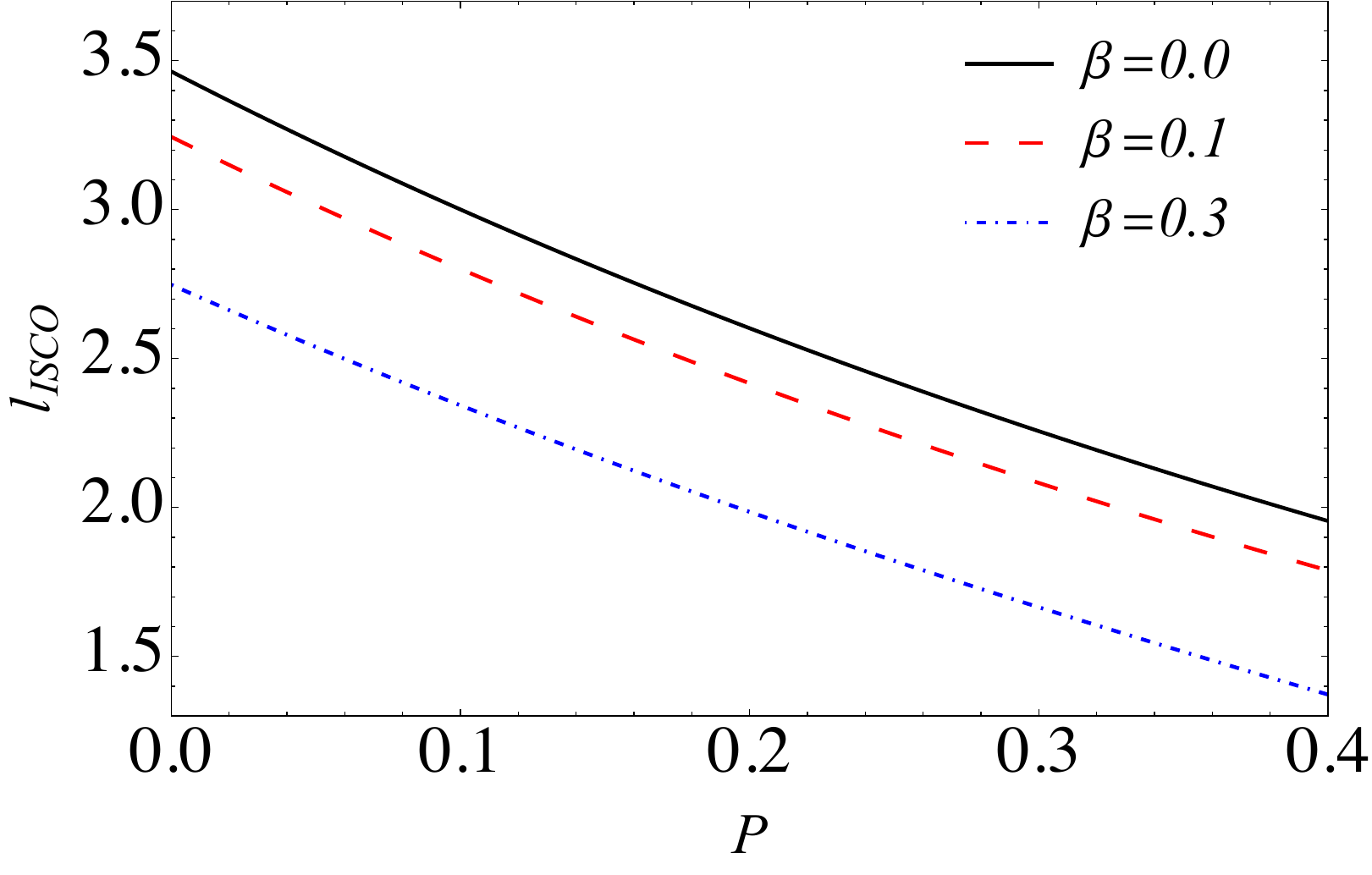}
			\includegraphics[width=0.32\textwidth]{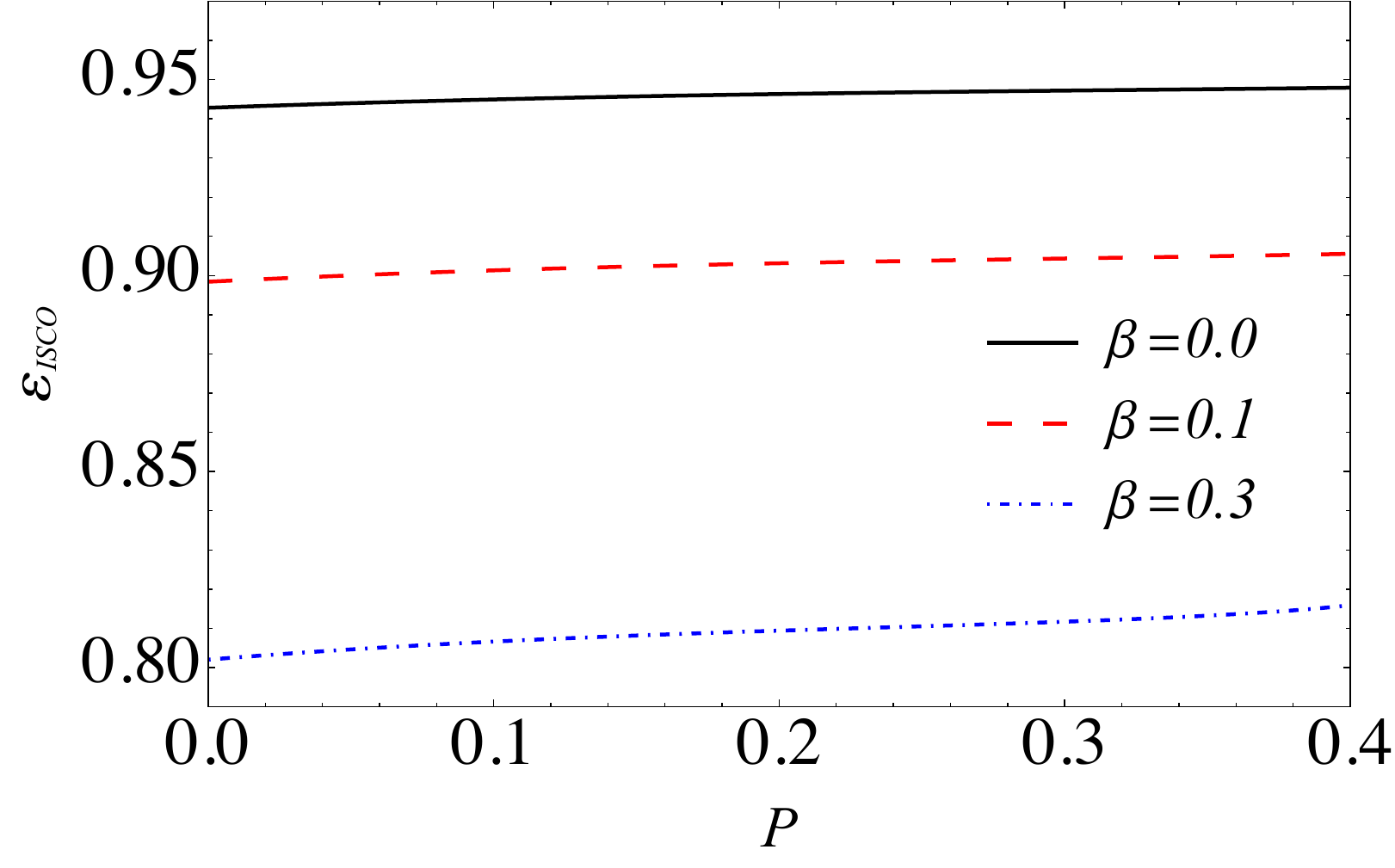}
			\caption{\label{ISCO magnetized} The ISCO parameters, $r_{ISCO}$, $l_{ISCO}$, $\mathcal{E}_{ISCO}$ for the magnetized particles are plotted as a function of the quantum correction parameter $P$ for various combinations of the magnetic interaction parameter $\beta$.}
		\end{figure*}
		\begin{figure*}[!htb]
			\centering
			\includegraphics[width=1\textwidth]{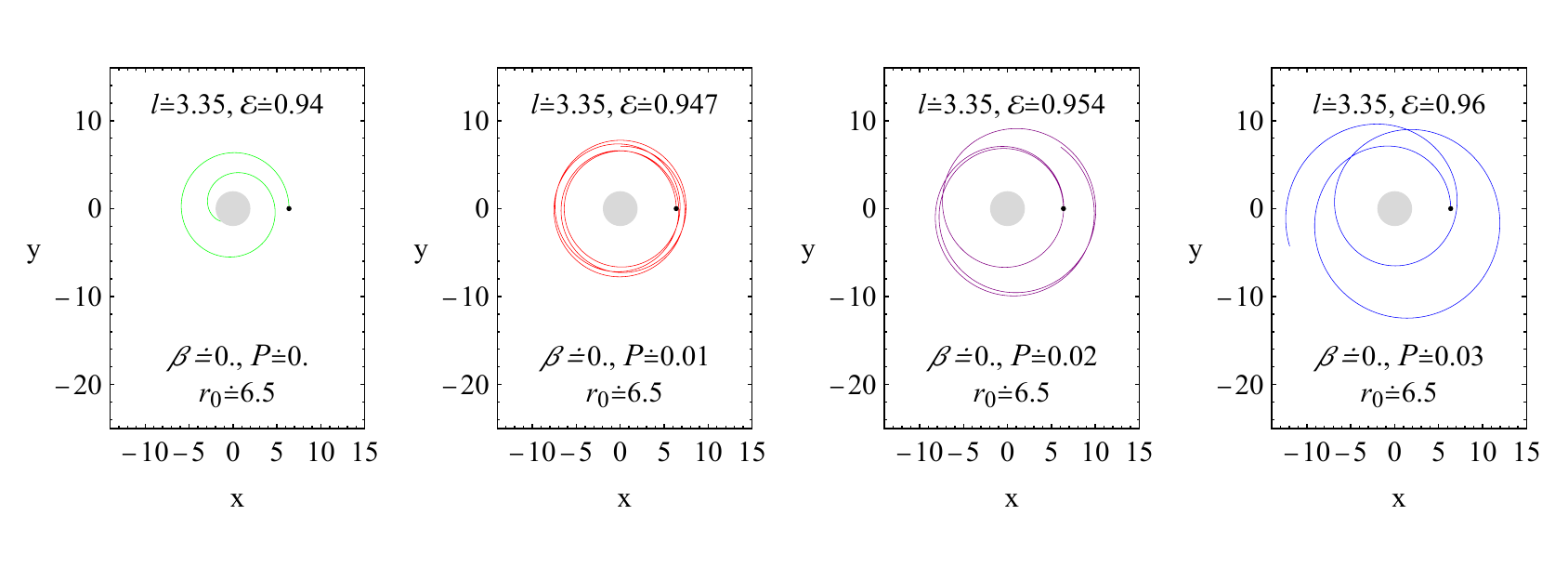}
			\includegraphics[width=1\textwidth]{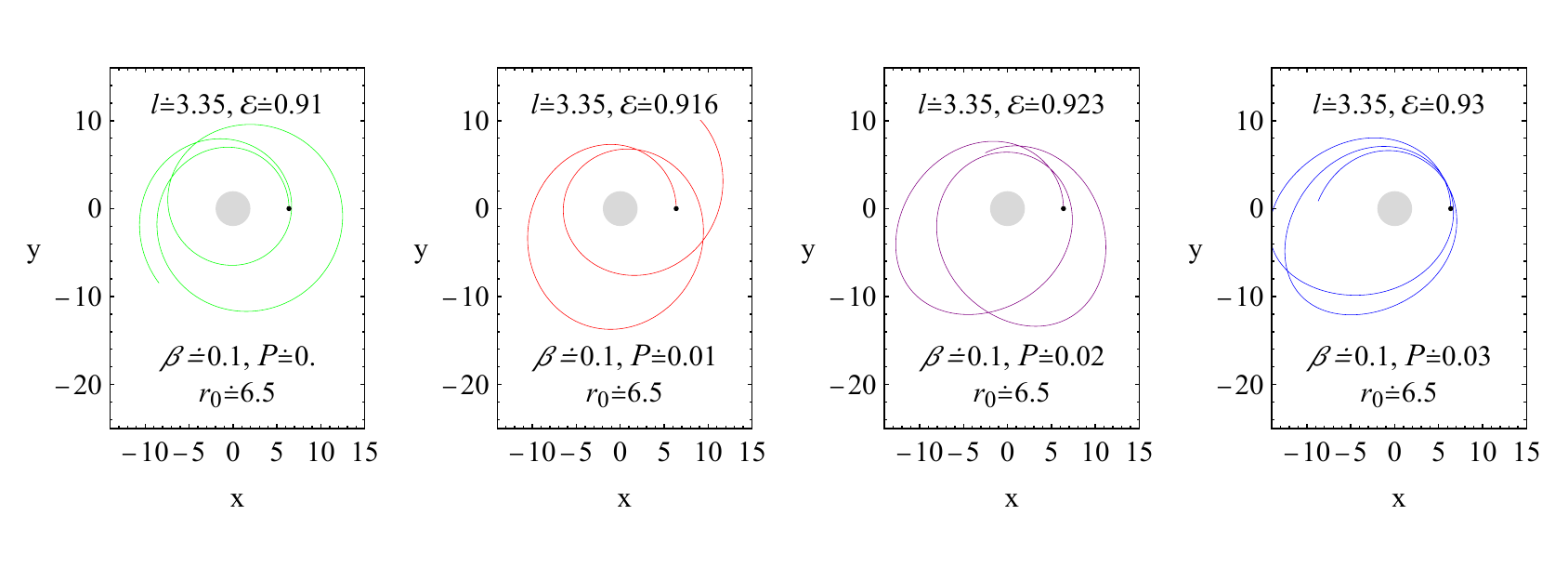}
			\caption{\label{fig:mag_par_trajectory} The trajectories of the test particle with magnetic dipole moment in the vicinity of the self-dual black hole immersed in an external uniform magnetic field for various possible cases. Here, we note that $\beta$ is the magnetic interaction parameter that represents the interaction between an external uniform magnetic field and a particle with a magnetic dipole moment.}
		\end{figure*}
		
		We then turn to analyze the effective potential for radial motion of the magnetized particles moving around the black hole in LQG in the case in which an external magnetic field is present. In Fig.~\ref{effective}, we depict the radial dependence of the effective potential $r$. Namely, we show the impact of the quantum correction parameter $P$ on the behavior of the effective potential for fixed magnetic interaction parameter $\beta$, while the impact of the magnetic interaction parameter for fixed $P$. As can be seen from Fig.~\ref{effective} the effective potential has two extreme points that correspond to stable and unstable circular orbits for particles with magnetic dipole moment in the LQG black hole vicinity, i.e., the minimum $r_{min}$ and the maximum $r_{max}$ represent stable and unstable circular orbits. From Fig.~\ref{effective}, the shape of the effective potential is shifted upward to its higher values, but the stable orbits are shifted to the right to larger $r$ as the quantum correction parameter $P$ increases. Similar behavior is also observed for unstable orbits. One can then deduce that the gravity in the background spacetime is significantly enhanced as a consequence of the effect of parameter $P$. 
		Similarly, we observe a resembling effect on the behavior of stable and unstable circular orbits under the effect of the magnetic interaction parameter $\beta$, which causes the radius of stable circular orbits to increase. Unlike stable orbits, the radius of unstable circular orbits begins to decrease as the magnetic coupling parameter $\beta$ rises. Additionally, it is important to find circular orbits on which particles with the magnetic dipole moment move around the self-dual black hole in LQG immersed in an external uniform magnetic field. To this end, one needs to impose the condition, $\frac{\partial^2V_{eff}}{\partial r^2}>0$ that solves to give stable circular orbits. We do however utilize a general condition $\frac{\partial V_{eff}}{\partial r}=\frac{\partial^2V_{eff}}{\partial r^2}=0$ to determine and analyze innermost stable circular orbits (ISCOs). In Fig.~\ref{ISCO magnetized}, we show the ISCO radius, angular momentum and energy as a function of the quantum correction parameter $P$ for various combinations of the magnetic interaction parameter $\beta$. As can be seen from the left panel of Fig.~\ref{ISCO magnetized}, the ISCO radius for the magnetized particles increases with the increasing magnetic interaction parameter, but it decreases under the effect of parameter $P$. Interestingly, it is observed that the effect of $P$ on the ISCO behavior reduces for larger values of the magnetic interaction parameter $\beta$. Furthermore, we also demonstrate other ISCO parameters, such as the angular momentum and the energy as a function of $P$ for different values of the magnetic coupling parameter $\beta$; see the middle and the right panels of Fig.~\ref{ISCO magnetized}. It is obvious that the combined effects of $\beta$ and $P$ can cause the angular momentum of the magnetized particles to decrease at the ISCO. However, the ISCO energy of magnetized particles slightly increases as the parameter $P$ increases, whereas it significantly decreases as a consequence of the magnetic interaction parameter $\beta$, as seen in the right panel of Fig.~\ref{ISCO magnetized}.  
		
		To gain a deeper understanding about the impact of the magnetic interaction and the quantum correction parameters, we also plot the magnetized particle trajectory restricted from moving on the equatorial plane of the self-dual black hole in LQG in the presence of an external magnetic field; see Fig.~\ref{fig:mag_par_trajectory}. There can exist various orbits, e.g., captured, bound and escape orbits due to different values of parameters, as seen in Fig.~\ref{fig:mag_par_trajectory}. From the particle trajectory, one can observe that the orbits are the captured when considering the vanishing both parameters, but these orbits turn to be bounded and then be escaping orbits when the quantum correction parameter $P$ increases; see the top row of Fig.~\ref{fig:mag_par_trajectory}. Interestingly, it is observed that the magnetized particle trajectories slightly become chaotic under the combined effects of $P$ and $\beta$, as shown in the bottom row of Fig.~\ref{fig:mag_par_trajectory}.                       
		
		Additionally, to be more informative we explore the ISCO parameters of the magnetized particle numerically and tabulate their values in Table~\ref{Table1}. Together with the ISCO radii, the specific angular momentum and energy of the magnetized particle moving at the ISCO, we also explore the magnetized particle's linear velocity (the azimuthal velocity) \cite{Shaymatov22a,Shaymatov2023} 
		\begin{eqnarray}\label{ISCO velocity}
			v_\phi=\Omega \sqrt{\frac{g_{\phi\phi}}{g_{tt}}},
		\end{eqnarray}
		with the Keplerian angular velocity/frequency, as tabulated in Table~\ref{Table1}, which is written by  
		\begin{eqnarray}
			\Omega=\sqrt{-\frac{\partial_rg_{tt}}{\partial_rg_{\phi\phi}}}\, .
		\end{eqnarray}
		\begin{table}[ht!]
			\centering
			\resizebox{.5\textwidth}{!}
			{
				\begin{tabular}{|l|c|c|c|c|c|r|}
					\hline
					$P$  & $\beta$ & ${r}_{ISCO}$ & $l_{ISCO}$ & $\mathcal{E}_{ISCO}$ & $v_{ISCO}$ & $\Omega_{ISCO}$ \\
					\hline
					0.00 & 0.0   & 6   & 2$\sqrt{3}$ & $\frac{2 \sqrt{2}}{3}$ &  1/2 & $\frac{1}{6 \sqrt{6}}$\\
					0.01 & 0.0   & 5.88230   & 3.36591 & 0.94385  & 0.49665 & 0.06916\\
					0.02 & 0.0   & 5.76906   & 3.27137 & 0.94491 & 0.49326 & 0.07027\\
					0.05 & 0.0   & 5.45429 & 3.00811 & 0.94807& 0.48293 & 0.07350 \\
					0.10 & 0.0   & 5.00308   & 2.62888 & 0.95335& 0.46514 & 0.07846 \\
					0.20 & 0.0   & 4.31952   & 2.04469 & 0.96365 & 0.42784 & 0.08627 \\ \hline
					&    &    &  &  &  &   \\
					0.00 & 0.1   & 6.10610   & 3.24455 & 0.89843 & 0.49349 & 0.06627 \\
					0.01 & 0.1   & 5.99407   & 3.15118 & 0.89959 & 0.48969 & 0.06723 \\
					0.02 & 0.1   & 5.88649   & 3.06118 & 0.90074 & 0.48584 & 0.06816 \\
					0.05 & 0.1   & 5.58893   & 2.81000 & 0.90409 & 0.47404 & 0.07080 \\
					0.10 & 0.1   & 5.16910   & 2.44653 & 0.90939 & 0.45349 & 0.07455 \\
					0.20 & 0.1   & 4.57350   & 1.88065 & 0.91868 & 0.40859 & 0.07859 \\
					\hline
				\end{tabular}
			}
			\caption{\label{Table1} Numerical values of the ISCO parameters, $r_{ISCO}$, $l_{ISCO}$, $\mathcal{E}_{ISCO}$, $v_{ISCO}$ and $\Omega_{ISCO}$  of the magnetized particles orbiting on the ISCO orbits around the self-dual black hole in the presence of an external magnetic field.} 
		\end{table} 
		The behavior of Figs.~\ref{ISCO magnetized} and \ref{fig:mag_par_trajectory} is also reflected in Table~\ref{Table1}. As can be seen from Table~\ref{Table1}, the particle's linear velocity at the ISCO decreases with the increasing quantum correction parameter $P$ and magnetic interaction parameter $\beta$, whereas the Keplerian angular velocity increases with an increasing $P$, but it decreases as a consequence of an increase in $\beta$. 
		
		We examined unique effects of the quantum correction parameter on the dynamics of particles with magnetic dipole moment and their non-geodesics around a self-dual black hole immersed in an external uniform magnetic field in LQG. We then focus on the dynamics of electrically charged particles around the black hole considered here. This is what we intend to take up in the next section. 
		
		\section{The motion of the charged particles near a self-dual black hole immersed in an external magnetic field in LQG}\label{Sec:elec_par}
		
		Here, we investigate the dynamics of electrically charged particles around a self-dual black hole immersed in an external uniform magnetic field in LQG. For that, we begin to write the Hamilton-Jacobi equation for electrically charged particle with the rest mass $m$ and charge $q$ as follows: 
		\begin{eqnarray}\label{HJ.Elec.}
			g^{\mu\nu}\left(\frac{\partial S}{\partial x^\mu}+qA_{\mu}\right)\left(\frac{\partial S}{\partial x^\nu}+qA_{\nu}\right)=-m^2,
		\end{eqnarray}
		where the four potential of electromagnetic field $A_\alpha$ is give in Eq.~(\ref{1.6}). The Lagrangian of the system for an electrically charged particle is then given by
		\begin{eqnarray}\label{Lagrangian.Electr}
			\mathcal{L}=\frac{1}{2}mg_{\mu\nu}u^{\mu}u^{\nu}+qu^\mu A_\mu\, ,
		\end{eqnarray}
		which allows to write the four-momentum as
		\begin{eqnarray}\label{4-momentum.Electr}
			p_\mu=mg_{\mu\nu}u^\nu+qA_\mu\, .
		\end{eqnarray}
		Following to the Hamilton-Jacobi action, $S=-Et+L\phi+S_r+S_\theta$, the equation of motion for a charged particle at the equatorial plane (i.e., $\theta=\pi/2$) is defined by 
		\begin{eqnarray}\label{Electr.motion2}
			g_{tt}\Big[1+g_{rr}\Dot{r}^2+\left(\frac{l}{r\sin{\theta}}+\frac{qA_\phi}{mr\sin{\theta}}\right)^2\Big]+\mathcal{E}^2=0\, ,\nonumber\\
		\end{eqnarray}
		where we have defined $\mathcal{E}=\frac{E}{m}$ and $l=\frac{L}{m}$ as the specific energy and angular momentum of the electrically charged particle per unit mass.
		Here, we note that for the further analysis we shall for simplicity restrict the motion of an electrically charged particle to the equatorial plane, $\theta=\frac{\pi}{2}$. Taking all into consideration the equation of the radial motion of charged particles moving around of the self-dual black hole is written as 
		
		\begin{eqnarray}\label{traj.electr.}
			\Dot{r}^2=\frac{1}{-g_{rr}g_{tt}}\Big(\mathcal{E}^2-V_{eff}\Big)\, .
		\end{eqnarray}
		From the radial equation of motion, the effective potential, $V_{eff}$, reads as
		\begin{eqnarray}\label{Veff.char.}
			V_{eff}=A(r)\left[1+\left(\frac{l}{r}+\frac{qA_\phi}{mr}\right)^2\right]\, .
		\end{eqnarray}
		\begin{figure*}[!htb]
			\centering
			\includegraphics[width=0.45\textwidth]{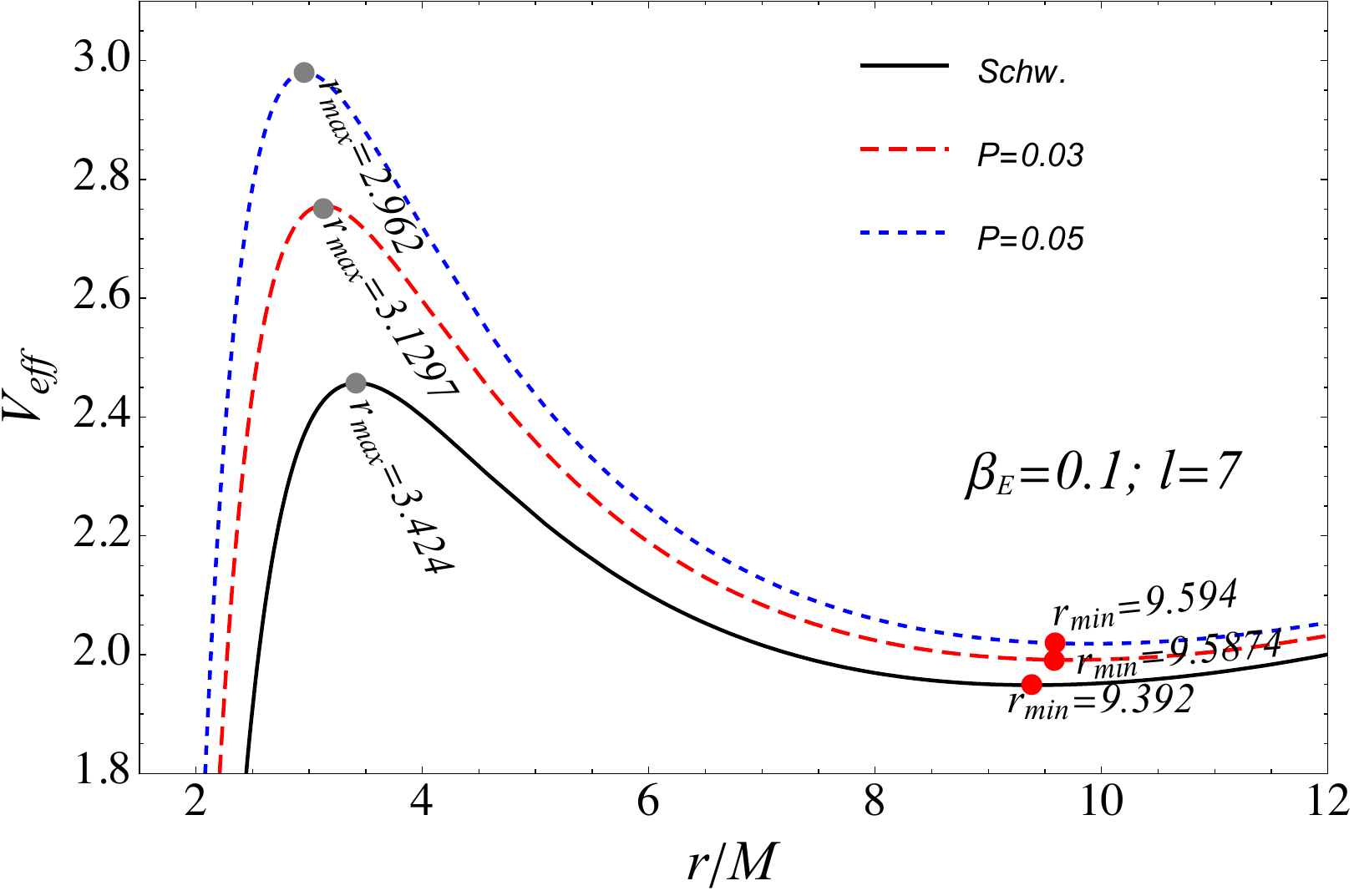}
			\includegraphics[width=0.45\textwidth]{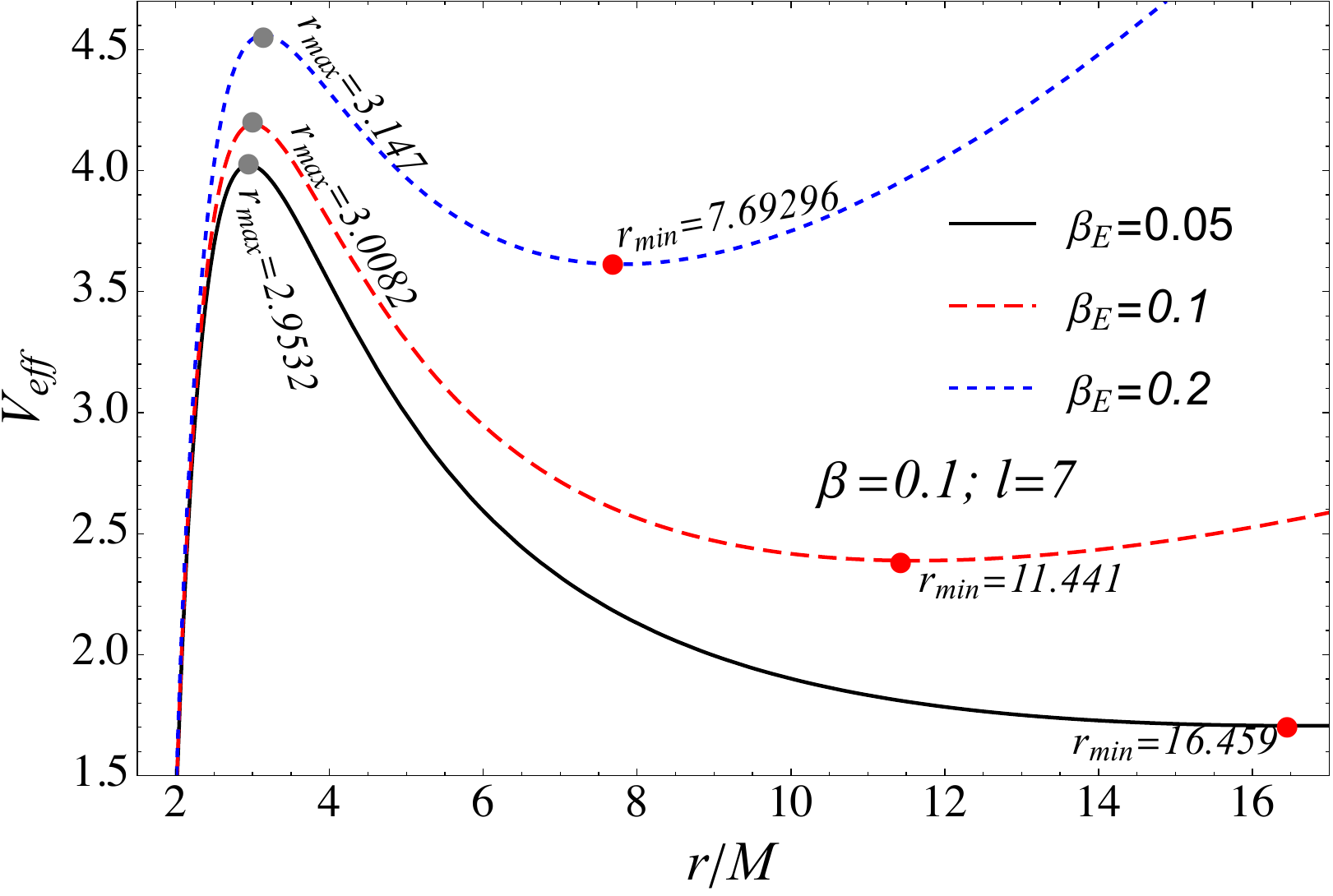}
			\caption{\label{Veff.ch} Radial profile of the effective potential for the electrically charged particles around the self-dual black hole in the presence of an external uniform magnetic field for various combinations of the quantum correction parameter $P$ (left panel) and the magnetic parameter $\beta_E$ (right panel). We note that the minima ($r_{min}$) and the maxima ($r_{max}$) correspond to the stable and unstable circular orbits, respectively.} 
		\end{figure*}
		\begin{figure*}[!htb]
			\includegraphics[width=1\textwidth]{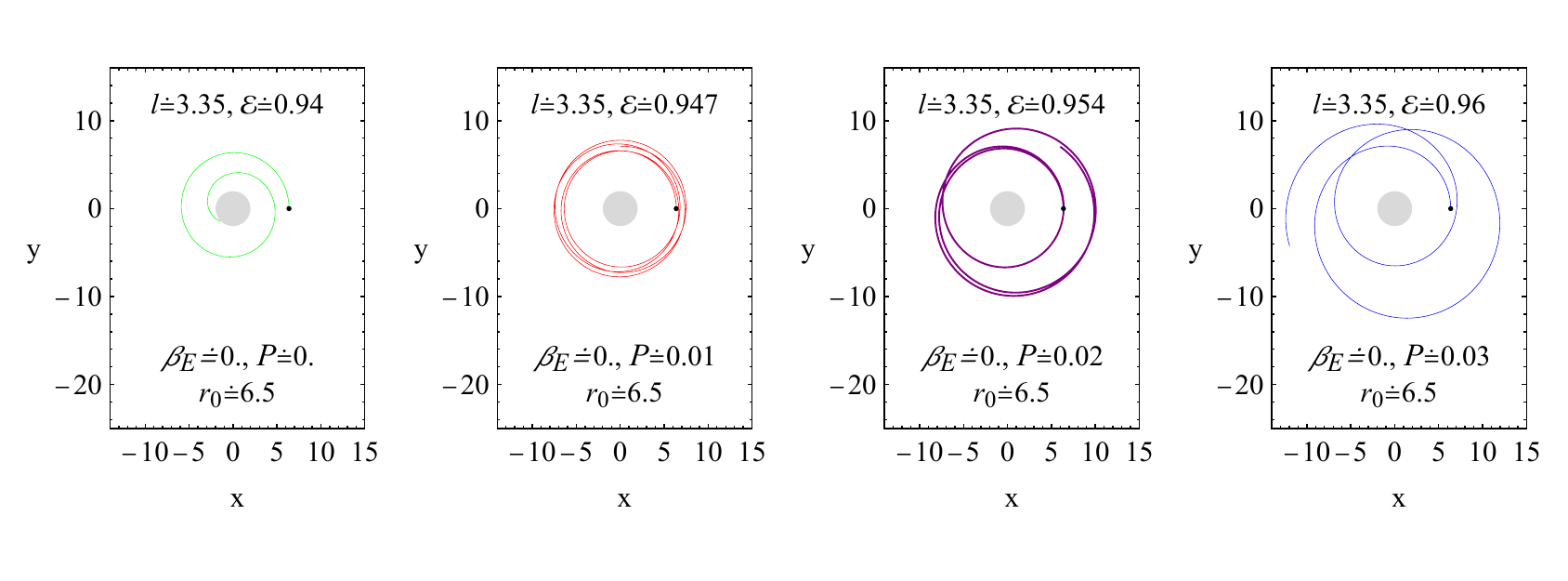}
			\includegraphics[width=1\textwidth]{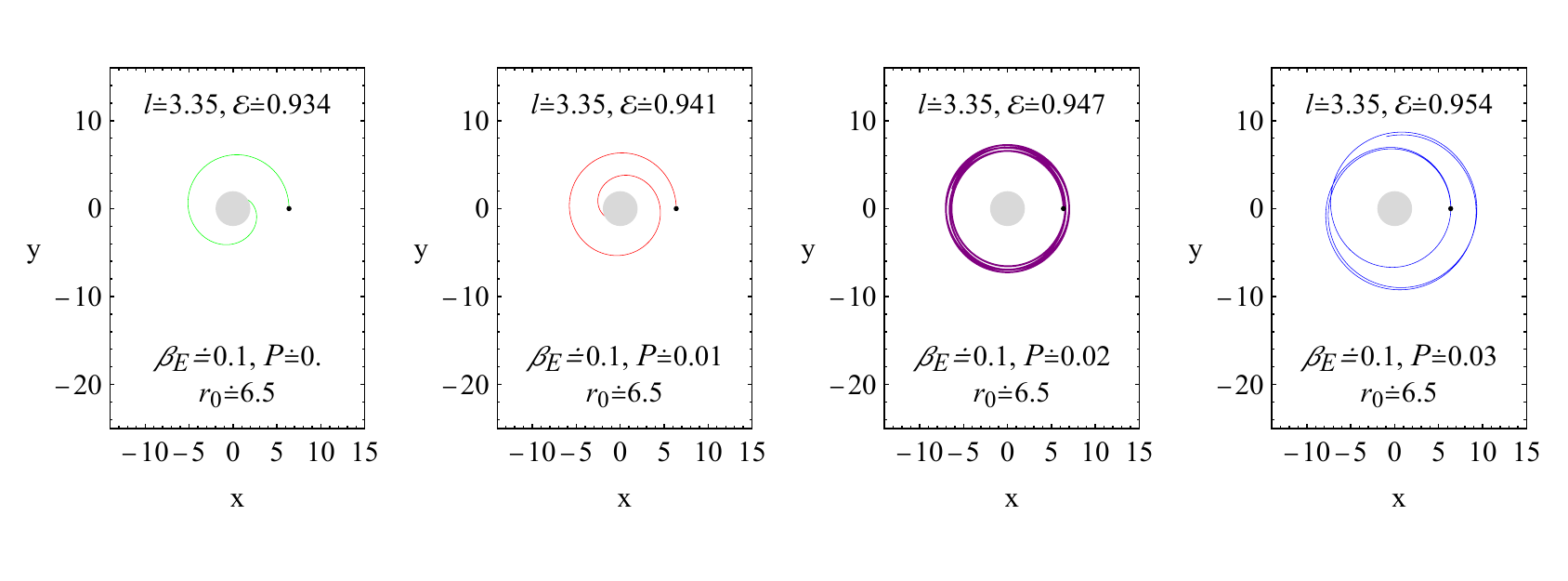}
			\caption{\label{fig:elec_par_trajectory} The trajectories of the electrically charged test particles in the vicinity of the self-dual black hole immersed in an external uniform magnetic field for various possible cases. Here, we note that $\beta_E$ is the magnetic parameter that represents the interaction between an external uniform magnetic field and an electrically charged particle.}
		\end{figure*}
		\begin{figure*}[!htb]
			\includegraphics[width=0.32\textwidth]{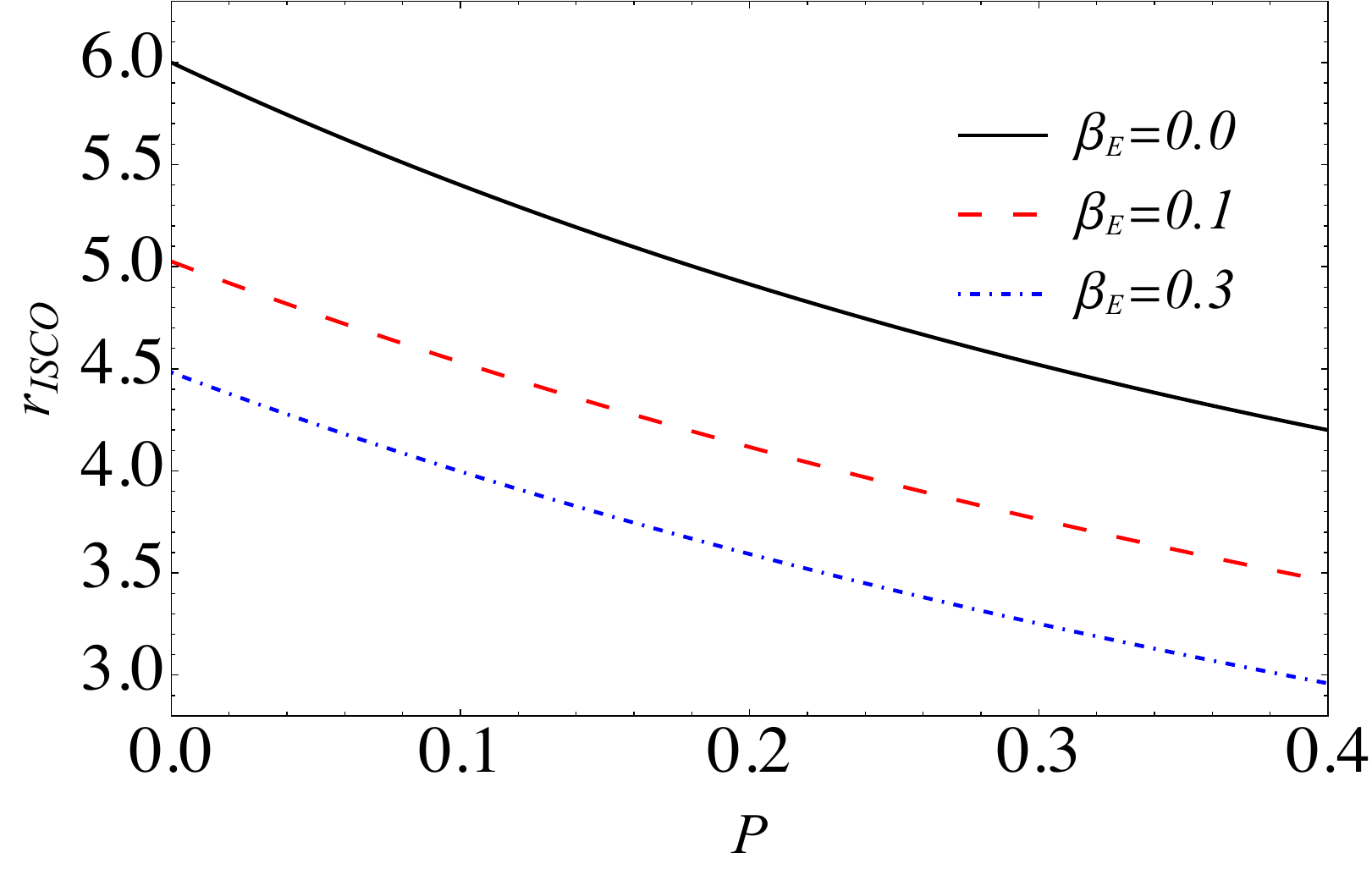}
			\includegraphics[width=0.32\textwidth]{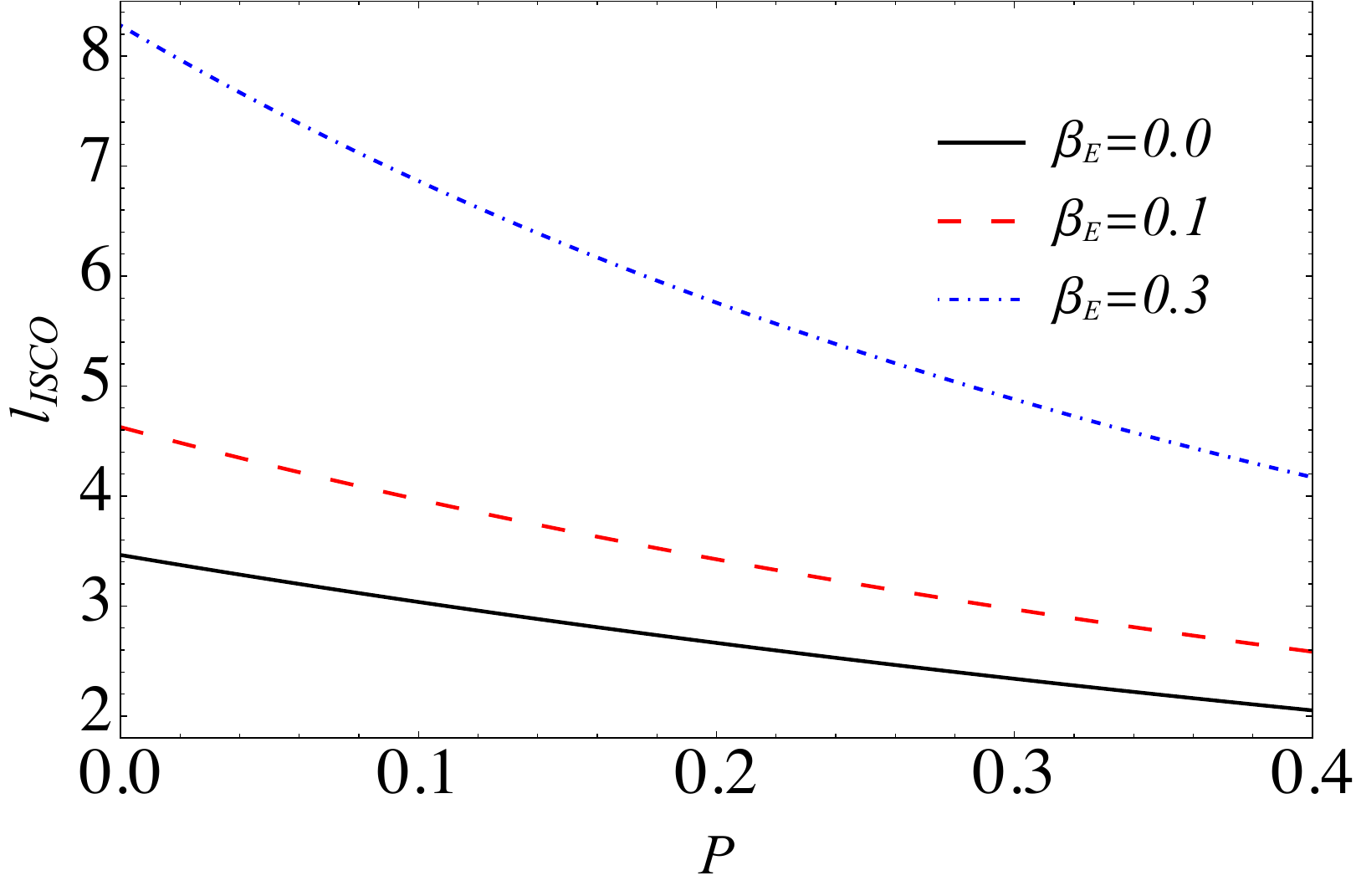}
			\includegraphics[width=0.32\textwidth]{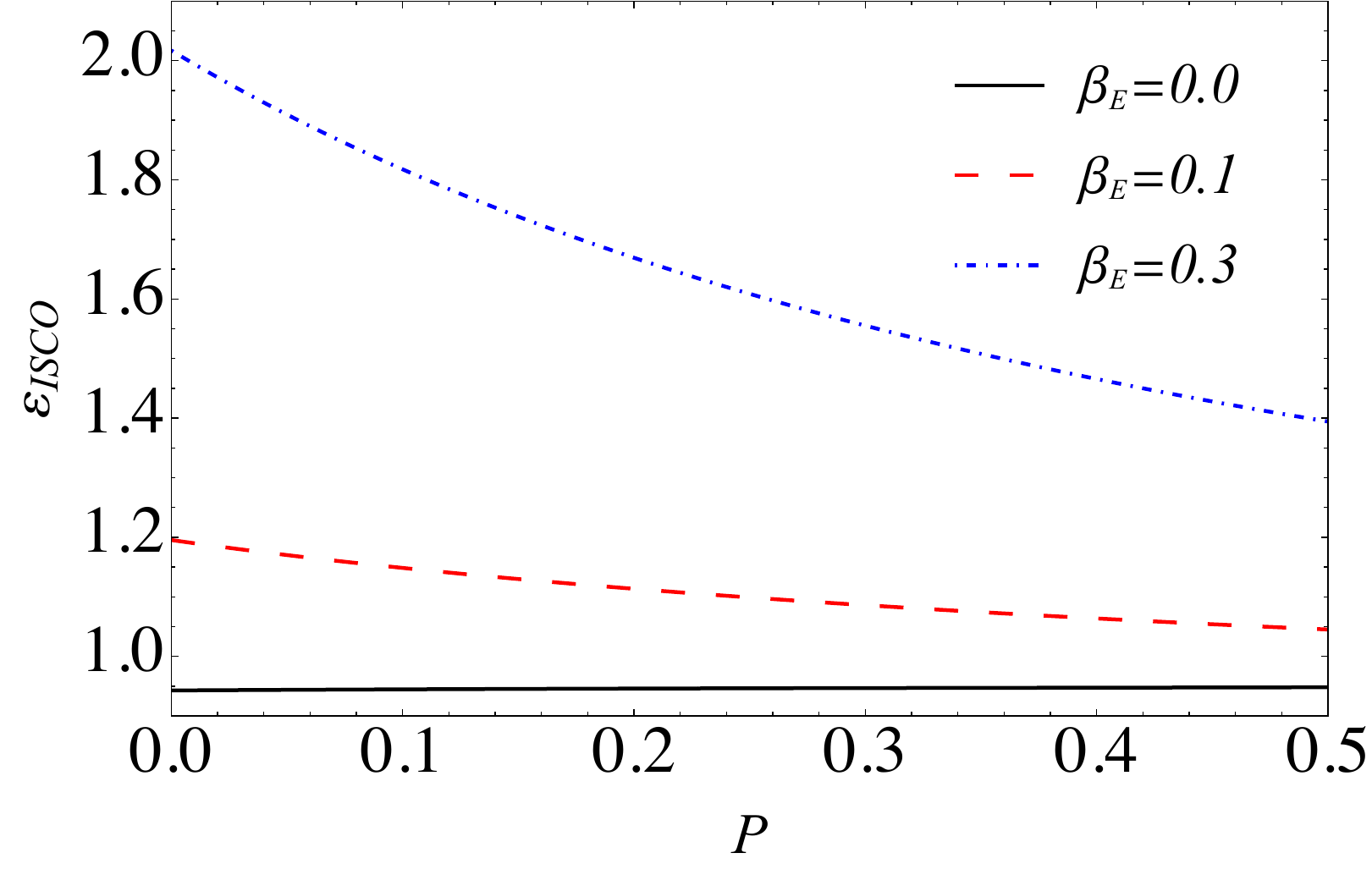}
			\caption{\label{ISCO charged} The ISCO parameters, $r_{ISCO}$, $l_{ISCO}$, $\mathcal{E}_{ISCO}$ for electrically charged particles are plotted as a function of the quantum correction parameter $P$ for various combinations of the magnetic parameter $\beta_E$.}
		\end{figure*}
		
		To gain a deeper understanding of the non-geodesic motion of an electrically charged particle around the black hole we shall further examine the effective potential for radial motion. In Fig.~\ref{Veff.ch}, we show the radial profile of the effective potential for the charged particle for various possible cases. The left panel of Fig.~\ref{Veff.ch} reflects the impact of the quantum correction parameter $P$ on the radial profile of the effective potential while keeping the magnetic interaction parameter $\beta$ fixed, while the right panel reflects the impact of $\beta_E$ for the fixed $P$. It is worth noting that the magnetic field can drastically influence the motion of a charged particle, causing its geodesics to alter around a black hole. Here, it is valuable to note that the magnetic field parameter plays a crucial role in comparing the relative strength of the magnetic and gravitational fields acting on a charged particle's motion around the black hole. Unlike, in a flat spacetime
		where the contribution of gravity no longer exists, a test particle with an electric charge $q$ and rest mass $m$ can have a characteristic cyclotron frequency in a magnetic field with strength $B$ 
		\begin{equation}
			\Omega_c=\frac{|qB|}{m}\, ,
		\end{equation}
		which serves as the magnetic field parameter. However, in a curved spacetime, together with gravity, a charged particle can, in addition to the characteristic cyclotron frequency, have the Keplerian frequency when orbiting around a black hole with mass $M$, i.e.,  
		\begin{equation}
			\Omega=\frac{r_g^{1/2}}{\sqrt{2}\, r^{3/2}}\, .
		\end{equation}
		where $r_g$ is the black hole gravitational radius $r_g$. The impact on the particle motion moving around a black hole (i.e., $r\sim r_g$) is interpreted by the dimensionless parameter, usually referred to as the magnetic field parameter $\beta_E$, which relates to the strength of the magnetic field (see, for example, \cite{Frolov10}) 
		\begin{eqnarray}\label{beta_E}
			\beta_{E}\equiv\frac{\Omega_c}{\Omega}\equiv\frac{qB_0M}{m}\, ,
		\end{eqnarray}
		where $B_0$ refers to the strength of the uniform external magnetic field surrounding the black hole. To be quantitative we estimate $\beta_{E}$ as a dimensionless quantity. For example, for a proton $\beta_{E}$ can take values on the order of $\sim 10^{7}$ around a black hole of mass $M\sim 10M_{\odot}$ and of $\sim10^{11}$ around a black hole of mass $M\sim10^9 M_{\odot}$ with the mass of the Sun, $M_{\odot}$. However, one can find that $\beta_{E}$ takes even larger values for an electron, showing that the magnetic field can drastically affect a charged particle motion due to the Lorentz force dominating over the gravitational field in the close vicinity of a black hole~\cite{Frolov10,Frolov12,Shaymatov21pdu}. As shown in Fig.~\ref{Veff.ch}, the curves of the effective potential shift upward toward larger values with increasing parameters $P$ and $\beta_E$, causing the strength of the gravitational potential to decrease. It is also observed from Fig.~\ref{Veff.ch} that an increase in the quantum correction parameter $P$ shifts stable circular orbits to larger $r$, while they become closer to the central object under the influence of the magnetic parameter $\beta_{E}$, resulting in the radius of stable circular orbits decreasing. Additionally, we demonstrate the trajectory of electrically charged particles restricted from moving on the equatorial plane of the self-dual black hole in LQG for various possible cases; see Fig.~\ref{fig:elec_par_trajectory}. Here, we analyze the effects of both $P$ and $\beta_{E}$ on the behavior of captured, bound, and escape orbits around the black hole. In Fig.~\ref{fig:elec_par_trajectory}, the top row shows the impact of parameter $P$ on the trajectory of electrically charged particles, while the bottom row shows the change in the behavior of particle trajectories when considering the magnetic parameter $\beta_{E}$. It is observed from the particle trajectory that orbits are captured when considering the vanishing of both parameters, but these orbits become bounded and escaping orbits when increasing the parameter $P$; as seen in the top row of Fig.~\ref{fig:elec_par_trajectory}. The interesting point to note here is that the role of the magnetic parameter can make bounded orbits turn into captured orbits, causing the electrically charged particle to fall into the black hole; see the bottom row of Fig.~\ref{fig:elec_par_trajectory}. Interestingly, one can observe that the lose of the electrically charged particle's energy, $\mathcal{E}$, decreases with an increasing quantum correction parameter $P$. This suggests that the effect of $P$ allows the particles to retain their energy when orbiting around the black hole, preventing them from falling into the black hole, as shown in the third panels of both the top and bottom rows of Fig.~\ref{fig:elec_par_trajectory}.
		
		Similar to what is observed to the previous section, here, we also turn to examine the ISCO parameters of the electrically charged particle moving around the self-dual black hole immersed in an external magnetic field in LQG. The following conditions, required for the particles to be on the ISCO,   
		\begin{eqnarray}\label{ISCO ch}
			V_{eff}(r)=\mathcal{E}_{ISCO}^2\,\,\mbox{and}\,\, \frac{\partial V_{eff}(r)}{\partial r}=\frac{\partial^2 V_{eff}(r)}{\partial r^2}=0\, . 
		\end{eqnarray}
		solve to give the ISCO parameters. In Fig.~\ref{ISCO charged}, we show the ISCO parameters, such as the ISCO radii, the energy and the angular momentum of the electrically charged particles. As shown in Fig.~\ref{ISCO charged}, the ISCO parameters decrease with an increasing parameter $P$, similar to what is observed in previous section. However, these parameters are influenced significantly by the magnetic parameter $\beta_{E}$. Interestingly, unlike the magnetic interaction parameter $\beta$ between magnetic field and particle's magnetic dipole momentum, an increase in the magnetic parameter $\beta_{E}$ decreases the ISCO radii but increases the angular momentum and the energy of the electrically charged particle, as seen in Fig.~\ref{ISCO charged}. 
		\begin{table}[ht!]
			\centering
			\resizebox{.5\textwidth}{!}
			{
				\begin{tabular}{|l|c|c|c|c|c|r|}
					\hline
					$P$  & $\beta_E$ & ${r}_{ISCO}$ & $l_{ISCO}$ & $\mathcal{E}_{ISCO}$ & $v_{ISCO}$ & $\Omega_{ISCO}$ \\
					\hline
					0.00 & 0.1   & 5.02571   & 4.62678 & 1.19536 & 0.57489 & 0.08875 \\
					0.01 & 0.1   & 4.93044   & 4.46667 & 1.18750  & 0.57108 & 0.09025 \\
					0.02 & 0.1   & 4.83816   & 4.31457 & 1.18017  & 0.56730 & 0.09174 \\
					0.05 & 0.1   & 4.57811   & 3.90153 & 1.16099  & 0.55607 & 0.09620 \\
					0.10 & 0.1   & 4.19459   & 3.33319  & 1.13652  & 0.53779 & 0.10353 \\
					0.20 & 0.1   & 3.57576   & 2.51660 & 1.10660  & 0.50334 & 0.11771 \\ \hline
					&    &    &  &   &  &  \\
					0.00 & 0.2   & 4.63195   & 6.32663 & 1.57136 & 0.61639 & 0.10031 \\
					0.01 & 0.2   & 4.53872    & 6.07161  & 1.54979  & 0.61315 & 0.10225 \\
					0.02 & 0.2   & 4.44858    & 5.83139 & 1.52955  & 0.60991 & 0.10420 \\
					0.05 & 0.2   & 4.19539   & 5.18963 & 1.47584  & 0.60022 & 0.11005 \\
					0.10 & 0.2   & 3.82382   & 4.33350  & 1.40524  & 0.58431 & 0.11985 \\
					0.20 & 0.2   & 3.22674   & 3.16359 & 1.31271  & 0.55439 & 0.13961 \\
					\hline
				\end{tabular}
			}
			\caption{\label{Table2} Numerical values of the ISCO parameters, $r_{ISCO}$, $l_{ISCO}$, $\mathcal{E}_{ISCO}$, $v_{ISCO}$ and $\Omega_{ISCO}$  of the electrically charged particles orbiting on the ISCO orbits around the self-dual black hole in the presence of an external magnetic field in LQG, similar to what is observed in Table~\ref{Table1} for the magnetized particles.} 
		\end{table}
		Furthermore, together with $r_{ISCO}$, $l_{ISCO}$, and $\mathcal{E}_{ISCO}$ of electrically charge particles orbiting at the ISCO, we examine the charged particle's linear (orbital) velocity 
		\begin{eqnarray}\label{ISCO velocity}
			v_\phi=\Omega \sqrt{\frac{g_{\phi\phi}}{g_{tt}}},
		\end{eqnarray}
		with the Keplerian angular velocity/frequency, $\Omega$, which is defined by \cite{Shaymatov22c}  
		\begin{eqnarray}\label{Eq:kep}
			\Omega^2&=&\Bigg[{\Omega_{0}^2}-2{g_{tt}}\left(\frac{qA_{\phi,r}}{mg_{\phi\phi,r}}\right)^2\pm 
			\frac{2qA_{\phi,r}}{mg_{\phi\phi,r}} \times \nonumber\\
			&\times&\left\{-\Omega_{0}^2g_{tt} -\Omega_{0}^4g_{\phi\phi}+\left(\frac{qA_{\phi,r}\,g_{tt}}{mg_{\phi\phi,r}}\right)^2 \right\}^{1/2}\Bigg] \times \nonumber\\
			&\times& \left[1+4g_{\phi\phi}\left(\frac{qA_{\phi,r}}{mg_{\phi\phi,r}}\right)^2\right]^{-1}\, ,
		\end{eqnarray}
		which, in the limit of $q\to 0$, reduces to 
		\begin{eqnarray}
			\Omega_0=\sqrt{-\frac{\partial_rg_{tt}}{\partial_rg_{\phi\phi}}}\, .
		\end{eqnarray}
		All numerical estimations for the ISCO parameters together with $v_{ISCO}$, $\Omega_{ISCO}$ are also reflected in Table~\ref{Table2} for the electrically charged particles orbiting on ISCO around the self-dual black hole in LQG in the presence of an external magnetic field. As shown in Table~\ref{Table1}, the electrically charged particle's linear velocity at the ISCO decreases with the rise in the quantum correction parameter $P$, but it increases under the effect of the magnetic field parameter $\beta_{E}$, while the Keplerian angular velocity increases with increasing $P$ and $\beta_{E}$.
		
		\section{\label{Sec:accr_disk}
			Accretion disk radiation for a Self-dual black hole immersed in an external uniform magnetic field in LQG}
		
		\begin{figure}
			\centering
			\includegraphics[width=0.45\textwidth]{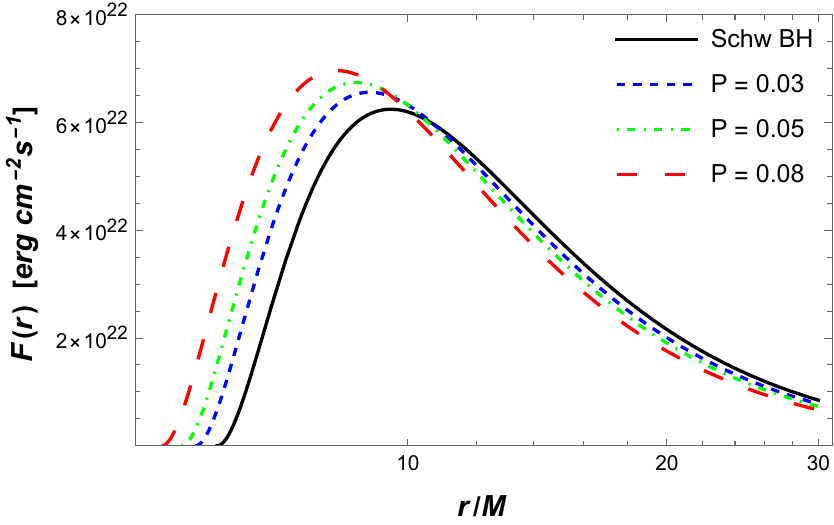}
			\caption{\label{fig:flux} The radial profile of accretions disk's electromagnetic radiation flux in the case of various possible cases associated with the quantum correction parameter $P$. 
			}
		\end{figure}
		\begin{figure}
			\centering
			\includegraphics[width=0.45\textwidth]{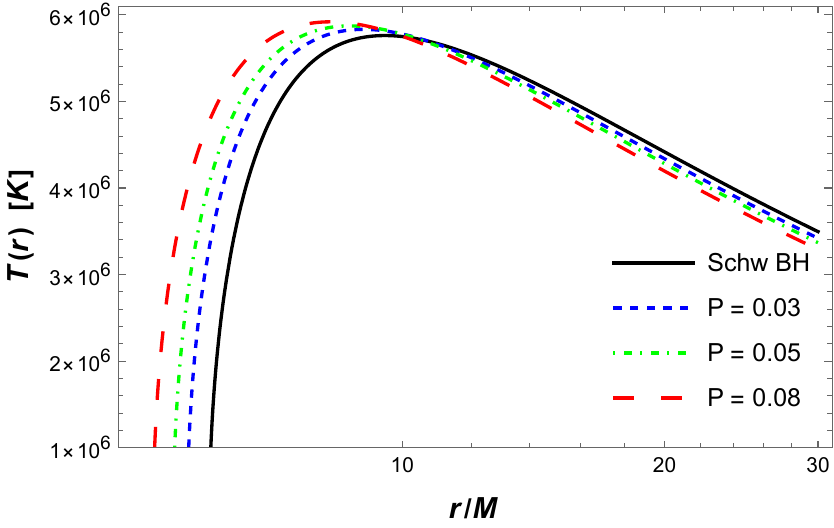}
			\caption{The radial profile of the accretions disk temperature for various values of quantum correction parameter $P$. }
			\label{temperature1}
		\end{figure}
		\begin{figure*}
			\centering
			\includegraphics[width=0.45\textwidth]{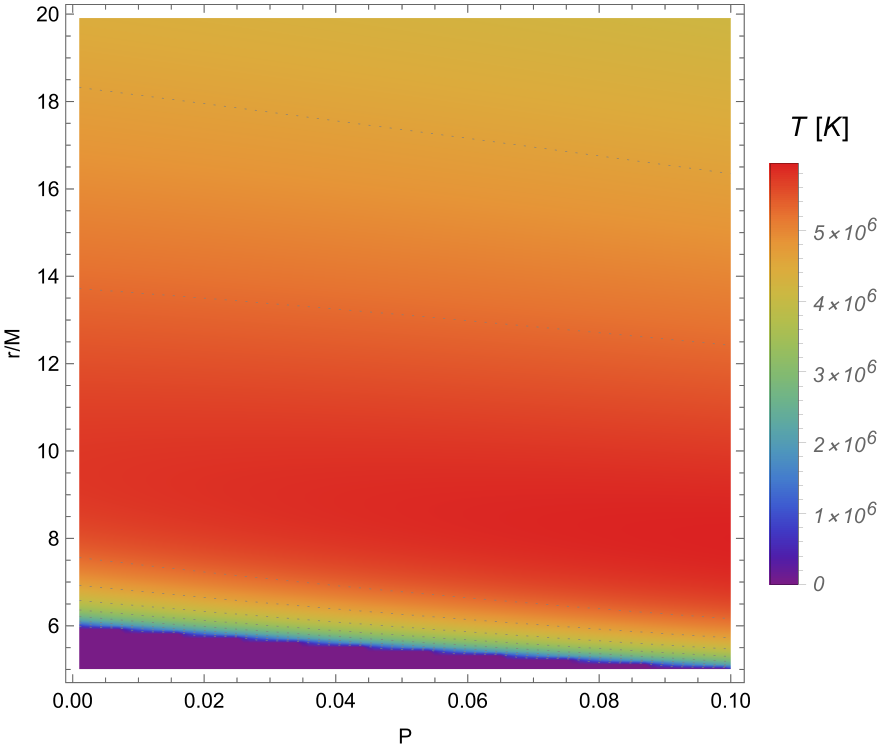}
			\includegraphics[width=0.45\textwidth]{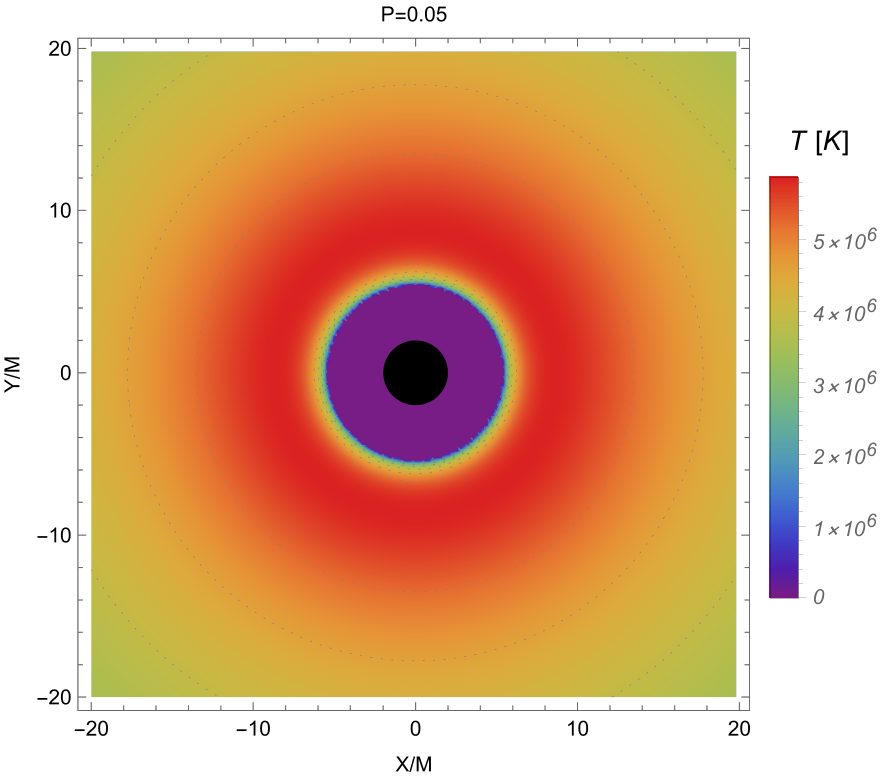}
			\caption{The accretion disk temperature profile is plotted under the influence of the quantum correction parameter $P$. The left and right panels respectively depict the parameter space of ($r,P$) and the density plot of the temperature profile of the accretion disk at the equatorial $X-Y$ plane of the self-dual black hole in LQG. Here, we denote $X$ and $Y$ as the Cartesian coordinates.}
			\label{temperature2}
		\end{figure*}
		\begin{figure}
			\centering
			\includegraphics[width=0.45\textwidth]{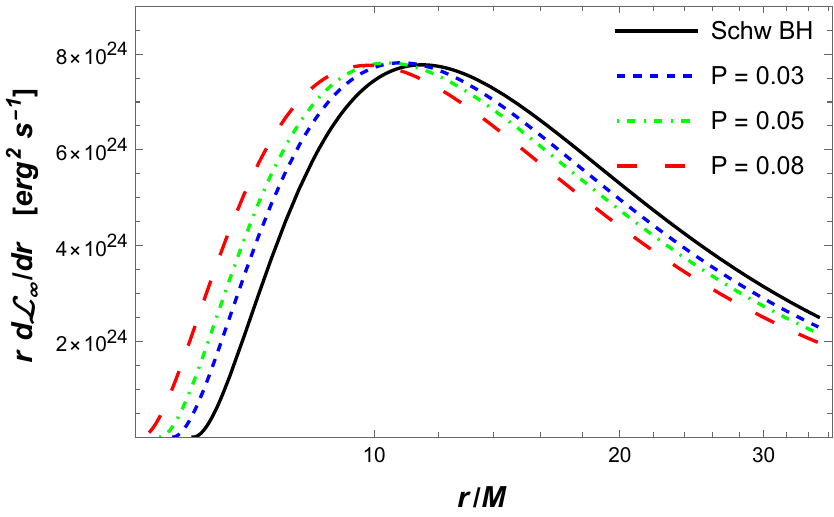}
			\caption{The radial profile of the differential luminosity of the accretion disk. 
			}
			\label{fig:luminosity}
		\end{figure}
		In this section, we investigate the radiation properties of the accretion disk composed of gas and dust orbiting around the self-dual black hole in LQG. In general, the accretion disk exists around compact objects like a black hole or a neutron star. According to the Novikov-Thorne model~\cite{Novikov:1973kta}, the accretion disk is optically thick and geometrically thin near a self-dual black hole in the LQG theory. The accretion disk is very thin vertically, but it spreads out significantly in the horizontal direction. As a result, the vertical size becomes insignificant compared to its extensive horizontal spread. Thus, the height $h$ is significantly smaller than the radius of the disk $r$, which extends horizontally, that is $h \ll r$. Based on astrophysical observations, the accretion disk can generally be characterized by the flux of electromagnetic radiation, disk temperature, and luminosity.  Here, we begin to consider the flux of electromagnetic radiation using the following equation
		\begin{equation}\label{flux}
			\mathcal{F}(r)=-\dfrac{\dot{M_0}}{4 \pi \sqrt{g}}\dfrac{\Omega_{,r}}{(E-\Omega L)^2} \int _{r_{ISCO}}^r (E-\Omega L) L_{,r} d r\, , 
		\end{equation}
		where $g$ represents the determinant of the three-dimensional subspace with coordinates $(t,r,\phi)$, which can be determined by $\sqrt{g}=\sqrt{-g_{tt}g_{rr}g_{\phi\phi}}$. Actually, here we consider the determinant of the near equatorial plane metric~\cite{Bambi_2012}.  It is to be emphasized that the flux of electromagnetic radiation of the accretion disk, ${\cal F}(r)$, can also be characterized by the unknown quantity referring to the mass accretion rate of the disk, which, for simplicity, can be taken as $\dot{M}_0=1$.  
		Additionally, the flux of electromagnetic radiation is influenced by the energy $E$ and angular momentum $L$ of neutral particles, along with the Keplerian frequency, $\Omega=\left(-{g_{tt,r}}/{g_{\phi\phi,r}}\right)^{1/2}$. {It should be noted that, for further analysis we shall, for simplicity, consider neutral particles in order to examine the effect of the quantum correction parameter on the radiation properties of a self-dual black hole in LQG.} It is complicated to derive the analytical expression for the flux of the accretion disk. Therefore, we resort to the numerical evaluation of ${\cal F}(r)$. In Fig.~\ref{fig:flux}, we demonstrate the radial profile of the accretion disk's electromagnetic radiation flux for different values of the quantum correction parameter $P$. It is clearly observed from Fig.~\ref{fig:flux} that the shape of the flux of the electromagnetic radiation shifts upward towards larger values, but it shifts downward towards smaller values at larger distances from the black hole as a result of an increase in the value of the quantum correction parameter, compared to the Schwarzschild black hole case (i.e., $P=0$) in Einstein gravity. 
		We further consider the Stefan-Boltzmann law to examine the temperature of the accretion disk, according to which the flux of the black body radiation can be defined by $\mathcal{F}(r)=\sigma T^4$, where $\sigma$ denotes the Stefan-Boltzmann constant.  
		We show the radial dependence of the accretion disk temperature around the black hole for various possible values of $P$ in Fig.~\ref{temperature1}. Similarly, the curves of the accretion disk temperature shift upward towards possibly larger values at distances close to the black hole with an increasing parameter $P$.  
		To be more detailed, we show a ``density plot" of the accretion disk temperature in Fig.~\ref{temperature2}. The left panel of Fig.~\ref{temperature2} reflects the change in the disk temperature with the change in the quantum correction parameter. Notably, it can be seen from the left panel that the accretion disk temperature is hot enough, especially close the black hole, but it starts to become cold towards to its outer edge, as shown as the red regions with the maximum temperature in the right panel of Fig.~\ref{temperature2}. 
		
		In addition to the accretion disk's flux and temperature, we also consider the differential luminosity as one of the important properties of the accretion disk, which can help to provide a deeper understanding of the disk radiation ~\cite{Novikov:1973kta,Shakura:1972te,Thorne:1974ve}. We then turn to the differential luminosity, which is defined by the following equation 
		\begin{equation}
			\dfrac{d \mathcal{L}_{\infty}}{d \ln{r}}=4 \pi r \sqrt{g} E \mathcal{F}(r)\, .
		\end{equation}
		Here, we shall for simplicity assume that the black body radiation can be considered to describe the radiation emission. To this end, we further define the spectral luminosity $\mathcal{L}_{\nu,\infty}$ which can be considered in terms of the radiation frequency (i.e., $\nu$) at infinity ~\cite{Boshkayev:2020kle,sym14091765,Shaymatov2023,Alloqulov_2024,Boshkayev_2021,boshkayev2023luminosityaccretiondisksrotating,kurmanov2024accretiondiskspropertiesregular,Alloqulov24EPJP} 
		
		\begin{equation}\label{luminosity2}
			\nu \mathcal{L}_{\nu,\infty}=\dfrac{60}{\pi^3} \int_{r_{ISCO}}^{\infty} \dfrac{\sqrt{g} E}{M^2}\dfrac{(u^t y)^4}{\exp\Big[{\dfrac{u^t y}{(M^2 \mathcal{F})^{1/4}}}\Big]-1} dr\, ,
		\end{equation}
		where $y=h \nu /k T_{\star}$, $k$ and $h$ are the constant of Boltzmann and the Planck constant, respectively, together with the mass $M$. Here, $u^t$ is the time component of the four velocity (for details see~\cite{Joshi_2014}). We note that $T_{\star}$ is here the characteristic temperature associated with the Stefan-Boltzmann law. This relation takes the form as 
		\begin{eqnarray}
			\sigma T_{\star}= \dfrac{\dot{M}_0}{4 \pi M^2}\, , 
		\end{eqnarray}
		where $\sigma$ represents the Stefan-Boltzmann constant. Taken together, we now analyze the differential luminosity of the accretion disk and demonstrate its radial profile in Fig.~\ref{fig:luminosity}. As it is expected, similar to the accretion disk flux, the differential luminosity initially increases and then decreases depending on distance, with an increasing quantum correction parameter $P$. Interestingly, it is observed that the differential luminosity can fall off at larger distances away from the black hole. This can be interpreted that the parameter $P$ would have a significant impact only in the close vicinity of the black hole. Additionally, we examine the spectral luminosity of the accretion disk around the self-dual black hole in LQG and display its frequency profile on a regular scale in Fig.~\ref{fig:spectrum}. As seen in Fig.~\ref{fig:spectrum}, the accretion disk's spectral luminosity is influenced by the impact of the quantum correction parameter, causing its curves to shift slightly downward towards smaller values.    
		\begin{figure}
			\centering
			\includegraphics[width=0.45\textwidth]{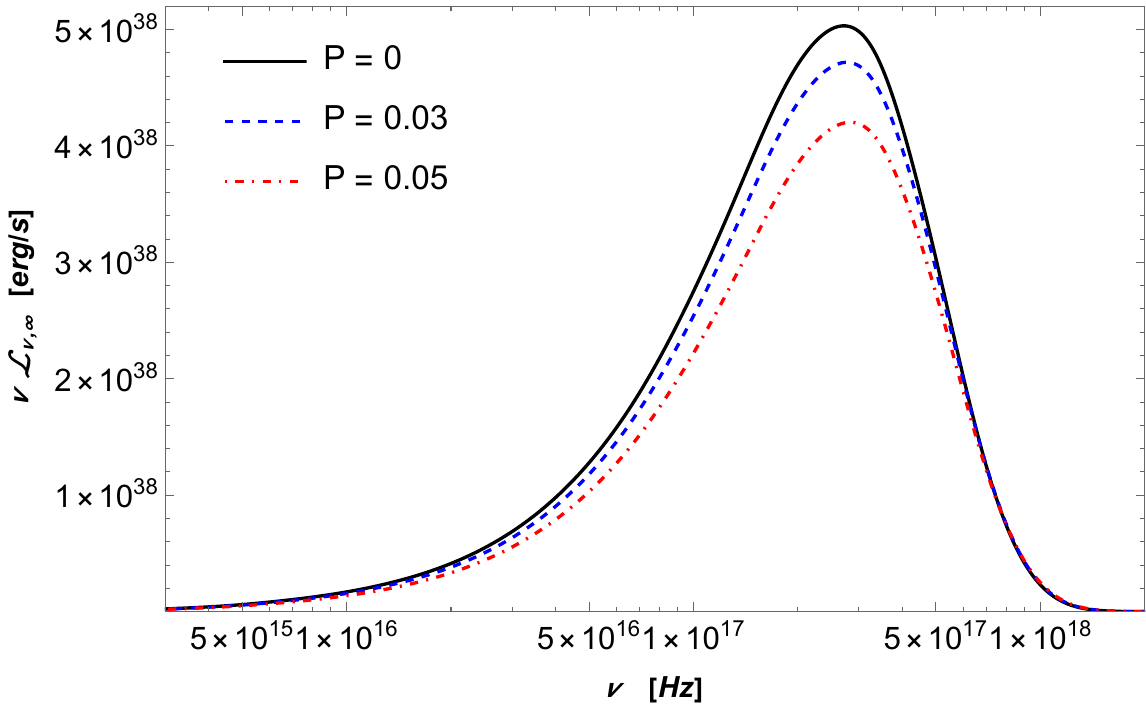}
			\caption{The accretion disk's spectral luminosity is plotted on a regular scale around a self-dual black hole in LQG for various values of the quantum correction parameter $P$. }
			\label{fig:spectrum}
		\end{figure}

		\section{Conclusions}\label{Sec:Con}
		
		In this paper, we considered a self-dual black hole immersed in an external asymptotically uniform magnetic field in LQG. To have information about spacetime geometry and test its nature one needs to delve into the effects of a quantum correction parameter and electromagnetic interactions on the particle geodesics. To this end, we studied the motion of magnetic dipoles and electrically charged particles in the environment surrounding the spacetime geometry in LQG. We derived the field equations and determined the electromagnetic four-vector potential for the case of the self-dual black hole in LQG. 
		
		We began to study the motion of magnetized particles around the black hole in the presence of an external magnetic field. We studied the stable and unstable circular orbits for particles with a magnetic dipole moment and showed that the stable orbits are shifted to exist at larger $r$ under the influence of the quantum correction parameter $P$. We also showed that similar behavior is also observed for unstable orbits. It was found that the gravity in the spacetime of the self-dual black hole in LQG is significantly enhanced by the impact of parameter $P$. Similarly, we also observed a similar behavior of stable and unstable circular orbits under the influence of the magnetic interaction parameter $\beta$. We also showed that stable orbits of electrically charged particles become closer to the black hole under the influence of the magnetic parameter $\beta_{E}$, causing the radius of these orbits to decrease. 
		
		We also studied the combined effects of the quantum correction parameter $P$ and the magnetic interaction parameter $\beta$ on the ISCO radius. We showed that the ISCO radius for magnetized particles decreases under the influence of the quantum correction parameter but it increases with an increasing magnetic interaction parameter $\beta$. However, unlike the magnetic interaction parameter $\beta$, the ISCO radius of electrically charged particles decreases with an increasing magnetic parameter $\beta_{E}$ decreases. Notably, we showed that the ISCO radius of magnetized particles is greater than that of electrically charged particles due to the magnetic field interaction. Further, we examined the ISCO parameters, such as the angular momentum $l_{ISCO}$ and the energy $\mathcal{E}_{ISCO}$, and showed that the angular momentum of the magnetized particles orbiting on the ISCO decreases under the effects of both the magnetic coupling parameter $\beta$ and the quantum correction parameter. Interestingly, we observed that the ISCO energy of magnetized particles increases due to the impact of $P$, while it decreases as a consequence of the influence of $\beta$.
		Unlike the magnetic interaction parameter $\beta$, the angular momentum and the energy of the electrically charged particles increase with a rise in the value of the magnetic parameter $\beta_{E}$. 
		
		We also showed that the particle's linear velocity $v_{ISCO}$ decreases, but the Keplerian angular velocity $\Omega_{ISCO}$ increases as a consequence of the influence of the quantum correction parameter $P$. However, when the effect of the magnetic interaction parameter $\beta$ is included, $v_{ISCO}$ and $\Omega_{ISCO}$ for the magnetized particles decrease with a rise in the value of $\beta$, whereas these parameters increase with an increasing magnetic field parameter $\beta_{E}$ for the electrically charged particles orbiting on the ISCO around the self-dual black hole immersed in an external uniform magnetic field in LQG.  
		
		Furthermore, we analyzed captured, bound and escape orbits of magnetized and electrically charged particles restricted from moving on the equatorial plane of the self-dual black hole in LQG for various possible cases. We showed that the orbits of the magnetized particles are captured when considering the vanishing of the quantum correction parameter $P$ and the magnetic interaction parameter $\beta$, but they turn to be bounded and then become escaping orbits as a consequence of an increasing quantum correction parameter $P$. Interestingly, it turns out that the magnetized particle trajectories slightly become chaotic under the combined effects of $P$ and $\beta$. However, from the trajectories of electrically charged particles we showed that the role of the magnetic parameter $\beta_{E}$ can make bounded orbits turn into captured orbits, causing the electrically charged particle to fall into the black hole. Unlike the magnetized particle case, we observed that the loss of the electrically charged particle's energy (i.e., $\mathcal{E}$) decreases with an increasing quantum correction parameter $P$, causing the particles to retain their energy when orbiting around the black hole and preventing them from falling into the black hole. This is a remarkable and distinguishing feature of orbits of electrically charged particles compared to the magnetized particles around the self-dual black hole in LQG.

		Finally, we investigated the accretion disk around the self-dual
		black hole as a primary source of information associated with
		the surrounding spacetime geometry and its properties in LQG. To provide valuable insights into the quantum correction parameter resulting from LQG, we studied its influence on the accretion disk radiative properties, such as the flux of electromagnetic radiation, the temperature and differential luminosity of the disk. Interestingly, we found that the curves of these accretion disk radiation quantities initially shift upward towards larger values and then shift downward towards smaller values at larger distances as a result of the rise in the value of the quantum correction parameter $P$, compared to the Schwarzschild black hole case (i.e., $P=0$) in Einstein gravity. 
		Additionally, we showed that the quantum correction parameter shifts the profile of the accretion disk's spectral luminosity downward, leading to a slight decrease in its magnitude. Based on the theoretical results, we showed that the quantum correction parameter $P$ causes a rise in the rate of electromagnetic radiation emitted by the accretion disk at closer distances around the self-dual black hole in LQG.  
		
		\section*{Acknowledgments}

		We are grateful to the anonymous Referee for constructive suggestions and comments. This work is supported by the Zhejiang Provincial Natural Science Foundation of China under Grants No. LR21A050001 and No. LY20A050002, the National Natural Science Foundation of China under Grants No. 12275238, the National Key Research and Development Program of China under Grant No. 2020YFC2201503, and the Fundamental Research Funds for the Provincial Universities of Zhejiang in China under Grant No. RF-A2019015.

		\appendix
		
		\section{A uniform magnetic field around the self-dual black hole in LQG}\label{A}
		
		Here, we solve the differential Eq.~(\ref{1.5}) analytically. To this end, we expend it into powers of small $P\ll 1$ and obtain the second order differential equation as follows:
		\begin{eqnarray}
			&& C_\phi'' \left(4 P r \left(r^2-M r\right)+r^2 (r-2 M)\right)+C'_\phi \Big(2 P \left(8 r^2-6 M r\right)+ \nonumber \\
			&& +2 \left(2 r^2-3 M r\right)\Big)-8 M P C_\phi+\frac{1}{r}O(P)^2=0\, ,
		\end{eqnarray}
		which solves to give the following solution
		\begin{eqnarray}\nonumber
			C_\phi=K_2(-4 Pr-r)^{-\frac{\sqrt{1-2 P}}{\sqrt{2 P+1}}-1} [2M (2 P+1)]^{1-\frac{\sqrt{1-2 P}}{\sqrt{2 P+1}}}\Big[(4 Pr+r)^{\frac{2 \sqrt{1-2 P}}{\sqrt{2 P+1}}}
		\end{eqnarray}
		\begin{eqnarray}\nonumber
			\times_2F_1\left(a_0,b_0;c_0;z_0\right)+K_1(-2M (2 P+1))^{\frac{2 \sqrt{1-2 P}}{\sqrt{2 P+1}}}
		\end{eqnarray}
		\begin{eqnarray}
			\times_2F_1\left(2-a_0,4-b_0;2-c_0;z_0\right)\Big]\,.
		\end{eqnarray}
		Here $K_1$, $K_2$ and $_2F_1(a,b;c;z)=\Sigma_{n=0}^{\infty}\frac{(a)_n(b)_n}{(c)_n}\frac{z^n}{n!}$ are the integration constants and Hypergeometric function, respectively. Also we have introduced new variables as $a_0=\frac{\sqrt{1-2 P}}{\sqrt{2 P+1}}-1$, $b_0=\frac{\sqrt{1-2 P}}{\sqrt{2 P+1}}+2$, $c_0=\frac{2 \sqrt{1-2 P}}{\sqrt{2 P+1}}+1$, $z_0=\frac{4 P r+r}{4 P M+2 M}$. It is obvious from the above equation that the constant $K_1$ can easily lose its meaning in the limit of $P=0$ that restores Schwarzschild black hole case. Hence, recalling Eq.~(\ref{a2}), we derive the following expression for $C_\phi$
		\begin{eqnarray}\nonumber
			&&C_\phi=K_2 \left(2 P+\frac{1}{2}\right)^{\frac{\sqrt{1-2 P}}{\sqrt{2 P+1}}-1} (M (2 P+1))^{1-\frac{\sqrt{1-2 P}}{\sqrt{2 P+1}}} r^{\frac{\sqrt{1-2 P}}{\sqrt{2 P+1}}-1}\times \\\label{a2}
			&&
			_2F_1\left(a_0,b_0;c_0;z_0\right).
		\end{eqnarray}
		In the case when $P=0$ (i.e., Schwarzschild black hole case), we derive $C_\phi=\frac{B_0}{2}$, which allows one to find $K_2=\frac{B_0}{2}$ easily. Taken all together, for the self-dual black hole immersed in an external uniform magnetic field in LQG , the four potential of the electromagnetic field can be obtained as follows:  
		\begin{equation}
			A^\mu=\frac{B_0}{2} (\mathcal{B})^{\mathcal{A}} \times _2F_1\left(\mathcal{A},\mathcal{A}+3;2\mathcal{A}+3;\mathcal{B}\right)\xi_\phi^\mu\, ,
			\label{a3}
		\end{equation}
		where we have defined $\mathcal{A}=\sqrt{\frac{1-2P}{1+2P}}-1$ and $\mathcal{B}=\frac{r(4P+1)}{2M(2P+1)}$ as new variables.

		\bibliographystyle{apsrev4-2}  
		\bibliography{Ref}

	\end{document}